\documentclass[twocolumn]{aastex62}
\usepackage{graphicx}
\usepackage{subfigure}
\usepackage{epstopdf}
\usepackage{amsmath} 
\usepackage{amsmath}
\usepackage{bm}
\usepackage{graphicx}
\usepackage{multirow}
\usepackage{enumerate}
\usepackage{amsmath}
\usepackage{bm}
\usepackage{graphicx}
\usepackage{natbib}
\usepackage{color}
\hypersetup{colorlinks,
linkcolor=blue,
citecolor=blue,
filecolor=blue,
urlcolor=blue}

\usepackage{soul}

\usepackage{enumerate}
\usepackage{textcomp}

\usepackage{multirow}
\usepackage{enumerate}

\begin{document}
\title{Post-Newtonian magnetohydrodynamics}
\correspondingauthor{Mahmood Roshan}
\email{mroshan@um.ac.ir}
\author{Elham Nazari}
\affiliation{Department of Physics, Ferdowsi University of Mashhad, P.O. Box 1436, Mashhad, Iran}
\author{Mahmood Roshan}
\affiliation{Department of Physics, Ferdowsi University of Mashhad, P.O. Box 1436, Mashhad, Iran}
\begin{abstract}

In this paper, we derive the post-Newtonian equations of the ideal Magnetohydrodynamics. To do so, we use the modern approach to post-Newtonian theory, where the harmonic gauge is used instead of the standard post-Newtonian gauge, and find the post-Newtonian metric in the presence of the electromagnetic fields. We show that although the electric field does not contribute in the metric and curvature of the spacetime, the magnetic field appears in the time-time component of the metric. The appearance of the magnetic field, in principle, leads to new relativistic contributions to the magnetohydrodynamic governing equations. Therefore, using the post-Newtonian metric, we find the relativistic corrections to the magnetohydrodynamic equations up to the first post-Newtonian order.
In addition, as usage of this derivation, we obtain a complete set of equations by which the behavior of a self-gravitating plasma can be determined in post-Newtonian gravity.

\end{abstract}


\section{introduction}\label{introduction}

Astrophysical plasmas are often strongly magnetized, and to study their physical properties the relativist physics is required. In these systems, the magnetic field can be so strong that the dynamical evolution is dominated by electromagnetic fields. For example objects like Active Galactic Nuclei (AGN), Gamma Ray Bursts (GRB) and neutron stars, especially a sub-class called magnetars, are known as systems where strong gravity and strong magnetic fields exist. For instance, the energy density in magnetars can be comparable or even larger than the kinetic energy. It is necessary mentioning that magnetars endowed with the super-strong magnetic field of $10^{14}-10^{15}\, \text{G}$ (\citealp{olausen2014mcgill}). It is believed that this strong magnetic field is the key parameter behind explosive events in neutron stars (\citealp{gourgouliatos2018strongly}). 

Hydrodynamical instabilities like Kelvin-Helmholtz and magneto-rotational instability are key processes to produce such a strong magnetic field, for magneto-rotational instability see \cite{giacomazzo2013formation}.  Kelvin-Helmholtz instability occurs during the merger of two neutron stars (as a binary system), in the shear layer between neutron stars, and increases the magnetic field to $10^{17}\, \text{G}$ \citep{price2006producing}.
 
We remind that in general relativity (GR), electromagnetic fields, like any other matter field, appear as a source for the curvature of spacetime. Therefore, it is natural to expect general relativistic effects associated with the strong electromagnetic fields in the above-mentioned systems. In fact, the magnetic energy density, i.e., $B^2/2\mu_0$, for $B=10^{15}\,\text{G}$ is $3.98\times 10^{27}\, \text{J}\, \text{m}^{-3}$. This energy density is equivalent to the rest mass energy density associated with a mass density of order $\rho\simeq 4.4\times 10^{10}\, \text{kg}\, \text{m}^{-3}$. This mass density  even exceeds the neutron stars density in the crust. Consequently, it is obvious that in order to study such strongly magnetized systems in GR, it is necessary to take into account general relativistic effects associated with electromagnetic fields. In other words, we need general-relativistic magnetohydrodynamics (GRMHD) to describe these systems.

Moreover, both the highly nonlinear nature of GR field equations and coupling to an electromagnetic field make it almost impossible and very complicated to study systems under realistic conditions in an analytic way. Therefore, GRMHD simulations play a crucial role to study these systems. There are several codes solving these equations numerically (e.g., see \citealp{giacomazzo2007whiskymhd} and \citealp{gammie2003harm}).

Furthermore, in addition to the numerical approaches, there is a powerful method in the context of GR known as post-Newtonian (PN) theory to solve GR equations approximately. This approach adds general relativistic effects step by step in an iterative way. 
Historically, Einstein, Infeld, and Hoffmann first adopted this theory and then, in the 1960s, it was developed by Fock, Chandrasekhar, and others (see \citealp{chandrasekhar1965post,chandrasekhar1967post,chandrasekhar1969conservation}; \citealp{chandrasekhar1969Second}; \citealp{chandrasekhar1970212}).
This method provides a useful tool to study GR tests and more importantly the emission of gravitational waves.
By utilizing this method, the experimental foundations of GR are investigated in \cite{thorne1971theoreticali,will1971theoreticalii,will1971theoreticaliii}.
Also, the gravitational radiations from binary systems are studied in \cite{blanchet1989post,blanchet1989higher,blanchet1995gravitational}.
PN theory has been exploited in the recent discovery of gravitational waves to interpret the discovered signals, see \cite{abbott2017gw170817}.
As other applications of this method, we refer the reader to tests of GR in the solar system (\citealp{thorne1987300,will1994proceedings}) and the equation of motion of binary systems (\citealp{hulse1975deep,blandford1976arrival,epstein1977binary,damour1991orbital}).
This method is also used to study the gravitational radiation reaction (\citealp{chandrasekhar1970212,burke1971gravitational,blanchet2006}), the PN effects in accretion disk models around BHs (\citealp{demianski1997dynamics}), the PN dynamical instability of inspiraling neutron star (NS) binaries (\citealp{lai1996innermost,faber2000post}), the relativistic effects on Jeans analysis (\citealp{nazari2017post}), and the PN corrections of Toomre criterion (\citealp{kazemi2018post}). To survey the applications of this approximation, we refer the reader to \cite{will2014confrontation}.

It should be mentioned that unlike the PN version of hydrodynamics which has been comprehensively developed to include higher PN corrections (for a review on the subject see \citealp{blanchet2014gravitational,will2014confrontation}), unfortunately, not much attention has been given to PN theory in the presence of the electromagnetic field in the literature. To the best of our knowledge, there is only one paper which studies this issue, i.e., \cite{greenberg1971post}. Therefore the lack of comprehensive literature in post-Newtonian magnetohydrodynamics (MHD) along with the significance of the magnetic field in the astrophysical systems, motivated us to revisit this issue. We remind that magnetic field plays essential role in star formation, relativistic jets, accretion disks and Type II supernova explosions. Even complex phenomena in the surface of the Sun are driven by magnetic fields. Therefore we believe that more attention to the effects of the magnetic field in the relativistic situations is required when dealing with semi-analytic methods like PN theory.

On the other hand, \cite{greenberg1971post}, as the only paper in this field, uses the early version/classic approach to obtain the relativistic corrections to MHD up to the first PN order, hereafter 1\tiny PN \normalsize. As we will briefly discuss this approach in the following, there are some ambiguities in this approach. In fact the current PN community prefers the modern approach to the PN theory.

The classic approach is based on the standard formulation of the Einstein field equations. This method was first proposed by \cite{chandrasekhar1965post} to find the PN equations of hydrodynamics in GR. In this approach, PN theory is framed in the standard gauge. This gauge has been widely used in the older written works. 
In the classic method, there are some ambiguities in the integral domain of the PN potentials with non-compact sources. Consequently, we encounter divergent integrals. It is important to bear in mind that these ill-defined integrals only are solved by imposing ad hoc choices and justifications for the solutions.
Moreover, in this method, there are some limitations which makes in inefficient for studying high PN orders (\citealp{chandrasekhar1970212}). From this perspective, it also seems necessary to revisit the post-Newtonian MHD, and rewrite it in the context of the modern approach.

On the other hand, in the modern approach, PN theory is completely embedded within post-Minkowskian (PM) theory in which the Landau-Lifshitz formulation of the Einstein field equations is utilized and the harmonic gauge is adopted. In this method, the PN metric is restricted to the near-zone (we will introduce this zone in the following section). It is worth mentioning that, although the standard gauge was widely used in the older PN literature, the current PN community prefers the modern approach in which the harmonic gauge is fixed. The first reason is that by using this method, PN theory can be embedded in PM theory. 
In fact, the domain of the PN integrals automatically is restricted during the calculations in PM theory. So, one can see that the ambiguities of the ill-defined gravitational potentials are naturally resolved in the modern approach and consequently PN theory is based on a safe foundation.
The second reason is that at higher PN orders, where the boundary conditions are important and the PN potentials with non-compact sources play essential roles, the modern approach is more successful.
For a thorough review of the advantages of the modern approach compared to the classic approach, we refer the interested reader to \cite{poisson2014gravity}.

Therefore, in this paper, we focus our attention on the modern approach, and by utilizing this method, we obtain the metric of weakly curved spacetime within a PN charged system.
Then, by using this PN metric, we construct the secure foundations of MHD in 1\tiny PN \normalsize order.
In other words, we revisit the analysis presented in \cite{greenberg1971post}, and present a different analysis. First of all, we use the modern approach instead of the classic one. In this case, we derive the reliable MHD equations and the components of the spacetime metric in 1\tiny PN \normalsize approximation. More especially, we introduce a safe PN potential which is carefully constructed from the magnetic field as a source term extending over all space. We explicitly show that this potential is not divergent and does not suffer from the ambiguities in the classic approach presented in \cite{greenberg1971post}. Secondly, as the main advantage and final goal of this study, the self-gravitating plasma is studied in the (PN) MHD regime. In other words, we extract the MHD equations determining the behavior of PN plasma. We show that these equations are the minimum possible number of relations by which one can completely describe plasma in PN gravity.

It is worth mentioning that, using the results presented in this paper, one can easily follow the footprint of the retarded effects, especially those directly related to the relativistic effects of electromagnetic fields, in the PN metric. This fact would be more clear at higher PN orders where the retarded solutions, i.e., wave-zone potentials, have more contributions in the PN metric. In this case, this paper can be considered as the first step toward finding the higher order corrections.

This paper is organized as follows. We begin in Sec. \ref{Sec2} by obtaining the PN metric of the spacetime within the MHD fluid. In this section, the modern approach including the harmonic gauge is introduced. In Sec. \ref{MHD_PN gravity}, we derive the relativistic generalization of the MHD equations in PN gravity, i.e., the PN continuity equation of the mass density, the PN Maxwell equations, and the PN Euler equation. In Sec. \ref{PN plasma}, by utilizing these PN equations as well as the PN expansion of Ohm's law, we introduce the PN version of equations. Using these equations, one can completely determine the behavior of the self-gravitating plasma in PN gravity. Finally, in Sec. \ref{summery}, our results are discussed.

\section{Post-Newtonian metric in the presence of the electromagnetic fields }\label{Sec2}

In this section, we obtain the metric of a weakly curved spacetime within a quasi-neutral, collisional fluid. By quasi-neutral fluid, we mean that the system is not completely neutral, and one cannot totally ignore the electromagnetic fields and their associated effects. On the other hand, we assume that the system lies in the PN limit. In this case, velocities in the system are relatively small compared to the velocity of light (slow motion condition), and the gravitational fields are weak enough to ignore higher order PN effects. In other words, we restrict ourselves to 1\tiny PN \normalsize approximation. In fact, in order to have a realistic description of the relevant astrophysical systems, we assume that the above mentioned charged matter distribution is a gravitationally bound system.
Therefore, in addition to the weak field limit, the slow motion condition as a result of which all speeds within the matter distribution are much smaller than the speed of light, is also established. It is also necessary to assume that the electromagnetic fields are strong enough to appear in the metric of the spacetime, but are weak enough to retain the system in the PN regime.  
On the other hand, to describe stronger fields, naturally, one needs to increase the accuracy of the approximation by taking into account higher order PN contributions.

In this study, we restrict ourselves to 1\tiny PN \normalsize approximation
and employ the ideal MHD assumption to describe the collisional fluid in the modern approach.
In subsection \ref{sub2_tools}, we briefly review the Landau-Lifshitz formulation of the Einstein field equations, and assemble the main variables and formulas that are required to formulate the modern PN approximation. 
We also present the first and second iterations in subsection \ref{sub2_First}.
In subsection \ref{sub2_PN metric}, the PN metric of the MHD system is derived.
Finally, the PN geodesic equation is introduced in subsection \ref{PN_geod}.

\subsection{Assembling the variables and formulas}\label{sub2_tools}
We assume that the reader is familiar with the content of the modern approach. So, here, we quite briefly review the essential ingredients of this method. In the following when we need necessary and basic equations in modern approach, we refer to the corresponding equations in \cite{poisson2014gravity}, and avoid duplication.

In the Landau-Lifshitz formulation, an important tool is the gravitational potential tensor 
$h^{\alpha \beta}=\eta^{\alpha \beta}-\mathfrak{g}^{\alpha \beta}$, 
where $\eta^{\alpha \beta}=\text{diag}(-1,1,1,1)$ is the Minkowski metric and $\mathfrak{g}^{\alpha \beta}=\sqrt{-g}g^{\alpha \beta}$ is the metric tensor density,
which is built from the metric tensor $g^{\alpha \beta}$ and its determinant $g$.
Here, the Greek indices denote the four spacetime variables and run over the values $0, 1, 2, 3$.  
In this context, 
as we know, the appearance of the field equations of GR is as follows
\begin{equation}\label{wave_eq}
\Box h^{\alpha \beta}=-\frac{16\pi G}{c^4}\tau^{\alpha \beta}
\end{equation}
where $\Box=\eta^{\alpha \beta}\partial_{\alpha \beta}$ is the flat-spacetime wave operator and
$\tau^{\alpha \beta}=(-g)(T^{\alpha \beta}+t_{\text{LL}}^{\alpha \beta}+t_{\text{H}}^{\alpha \beta})$
is the effective energy-momentum pseudotensor 
built from the total energy-momentum tensor of the charged matter distribution $ T^{\alpha \beta}$, the Landau-Lifshitz pseudotensor $(-g)t_{\text{LL}}^{\alpha \beta} $, and the harmonic-gauge pseudotensor $(-g)t_{\text{H}}^{\alpha \beta}$ \footnote{The exact version of $t_{\text{LL}}^{\alpha \beta} $ and $t_{\text{H}}^{\alpha \beta}$  is presented in equations (6.5) and (6.53) in \cite{poisson2014gravity}.}.

The retarded solution to the equation (\ref{wave_eq}) is given by
\begin{equation}\label{general_int}
h^{\alpha\beta}\left(t,\bm{x}\right)=\frac{4G}{c^4}\int \frac{\tau^{\alpha\beta}\left(t-|\bm{x}-\bm{x}'|/c,\bm{x}'\right)}{|\bm{x}-\bm{x}'|}d^3\bm{x}'
\end{equation}
where $\bm{x}'$ is the source-point position, and $\bm{x}$ is the field-point position. The domain of this integration is extended over the past light cone of the field point. This light cone is decomposed into near-zone and wave-zone contributions by an arbitrary radius, $\mathcal{R}$, which is smaller than the characteristic wavelength of the gravitational radiation, $\lambda_c$, emitted by the matter distribution.
As previously mentioned, PN theory is an approximate method within the near-zone (i.e., when $|\bm{x}|<\mathcal{R}$) and works in a system where weak-field gravity and slow-motion condition are established.
Therefore, we should only evaluate the gravitational potentials in which $\bm{x}$ lies in the near-zone. In this case, however, the source point $\bm{x}'$ can be located at both near and wave-zones (i.e., $|\bm{x}'|<\mathcal{R}$ and $|\bm{x}'|>\mathcal{R}$ respectively). So one can particularly decompose the domain of the integration (\ref{general_int}) into near-zone and wave-zone pieces and rewrite this equation as
$h^{\alpha\beta}=h^{\alpha\beta}_{\mathcal{N}} +h^{\alpha\beta}_\mathcal{W}$,
where $h^{\alpha\beta}_{\mathcal{N}}$ and $h^{\alpha\beta}_{\mathcal{W}}$ are the near-zone and the wave-zone portions of the light-cone integral of equation (\ref{general_int}), respectively. The complete form of these contributions is introduced in chapter 7 of \cite{poisson2014gravity}.

Throughout this paper, we assume that the matter distribution has a perfect fluid component. So for this component we have  
\begin{equation}\label{T_fluid}
T^{\alpha\beta}_{\text{fluid}}=\left(\rho+\frac{\epsilon}{c^2}+\frac{p}{c^2}\right)u^{\alpha}u^{\beta}+pg^{\alpha\beta}
\end{equation}
where $\rho$ is the proper mass density, $\epsilon$ is the proper internal energy density, and $p$ is the pressure. Also, $u^{\alpha}$ is the velocity field expressed as $u^{\alpha}=\gamma\left(c,\bm{v}\right)$ in which $\bm{v}$ is the fluid's velocity field and the factor $\gamma$ is equal to $u^0/c$.
In this method, the rescaled mass density, $\rho^*=\sqrt{-g}\gamma\rho$, and the internal energy per unit mass, $\Pi=\epsilon/\rho^*$,  are introduced as two important matter variables in the PN theory.

Because there is an electromagnetic field in our system and this field can carry energy and momentum, the total energy-momentum tensor also includes the field's contribution $T^{\alpha\beta}_{\text{field}}$. Therefore $T^{\alpha\beta}=T^{\alpha\beta}_{\text{fluid}}+T^{\alpha\beta}_{\text{field}}$, in which
\begin{equation}\label{T_field}
T^{\alpha\beta}_{\text{field}}=\frac{1}{\mu_0}\left(F^{\alpha\mu}F^{\beta}_{\mu}-\frac{1}{4}g^{\alpha\beta}F_{\mu\nu}F^{\mu\nu}\right)
\end{equation}
includes the electromagnetic field tensor $F^{\alpha\beta}$ and the vacuum permeability $\mu_0$.
In the laboratory frame, the components of 
$F^{\alpha\beta}$ are known as
\begin{equation}\label{E&B}
F^{0j}=c^{-1}E^{j}
\quad\mathrm{,}\quad 
F^{ij}=\epsilon^{ijk}B_k
\end{equation}
where $\epsilon^{ijk}$ is the permutation symbol, $E^{j}$ is the electric field, $B^{j}$ is the magnetic field, and the Latin indices run over the values $x$, $y$, and $z$.
Indices on $F^{\alpha\beta}$ are lowered by the metric $g_{\alpha\beta}$.
Here, the electric field is imagined to be of the same order of magnitude as the magnetic field.

In this method, $\tau^{\alpha\beta}$ is conserved when the harmonic gauge condition, i.e., $\partial_{\beta}h^{\alpha \beta}=0 $, is satisfied. So we have  
\begin{equation}\label{conserv_eq}
\partial_{\beta}\tau^{\alpha\beta}=0
\end{equation}
Equations (\ref{wave_eq}) and (\ref{conserv_eq}) are the main expressions in this framework.

As we know, the general form of the components of the PN metric in terms of $h^{\alpha\beta}$ in 1\tiny PN  \normalsize context, where we need $g_{00}$ to order $c^{-4}$, $g_{0j}$ to order $c^{-3}$, and $g_{jk}$ to order $c^{-2}$, can be written as
\begin{eqnarray}
\label{g00}
g_{00} &=&-1+\frac{1}{2}h^{00}-\frac{3}{8}\left(h^{00}\right)^2+\frac{1}{2}h^{kk}+O(c^{-6})\\
\label{g0j}
g_{0j} &=& -h^{0j}+O(c^{-5})\\
\label{gjk}
g_{jk} &=& \delta_{jk}\left(1+\frac{1}{2}h^{00}\right)+O(c^{-4})\\
\label{det-g}
\left(-g\right) &=& 1+h^{00}+O(c^{-4})
\end{eqnarray}
where $\delta_{jk}$ is the Kronecker delta.
To evaluate the complete contribution of the potential components in the 1\tiny PN \normalsize correction of the metric, we actually have to carry out two iterations of the wave equation (\ref{wave_eq}). We do this in the subsequent subsections where the $n$th numerical index of the potentials expresses the $n$th iteration of the wave equation. In this iteration, the source term is $\tau_{n-1}^{\alpha\beta}$ which is constructed in the $\left(n-1\right)$th iteration. 
It should be also noted that the gauge condition/conservation statement is enforced in the second step of the iterative procedure.
In the following derivation, we drop the calculation related to the fluid contribution and only introduce the field part. For the fluid contribution one can find a detailed description in \cite{poisson2014gravity}.

\subsection{First and second iterations}\label{sub2_First}
In the first iteration, our final goal is to construct $g^{\alpha\beta}_1$ from $h^{\alpha\beta}_1$. To do so, 
we should derive the source term of equation (\ref{general_int}), i.e., $\tau^{\alpha\beta}_0$, in terms of $g^{\alpha\beta}_0$. 
Inserting $h^{\alpha\beta}_0=0$ within equations (\ref{g00})-(\ref{gjk}), we obtain that $g^{\alpha\beta}_0=\eta^{\alpha\beta}$. 
Consequently, 
one can easily show that the Landau-Lifshitz and harmonic pseudotensors vanish.
So we see that $\tau^{\alpha\beta}_0=T^{\alpha\beta}_0$; and to obtain the source term, we need only  know two pieces of $T^{\alpha\beta}_0$, i.e., $T^{\alpha\beta}_{0\,\text{fluid}}$ and $T^{\alpha\beta}_{0\,\text{field}}$.
By using equations (\ref{T_fluid}) and (\ref{T_field}), and 
replacing $g^{\alpha\beta}$ by $\eta^{\alpha\beta}$ in them, one can easily obtain the components of $T^{\alpha\beta}_{0\,\text{fluid}}$ and $T^{\alpha\beta}_{0\,\text{field}}$ in the Minkowski spacetime, respectively.
On the other hand, as we shall see in the following,  
in the second iteration of the field equations, we need only calculate $h^{00}_1$ to order $c^{-2}$ and $h^{0j}_1$ to order $c^{-3}$ and also discard $h^{jk}_1$.
This issue and the appearance of the equation (\ref{general_int}) remind us that the sufficient orders for the time-time, time-space, and space-space components of $T^{\alpha\beta}_0$ are $O(c^2)$, $O(c)$, and $O(1)$, respectively. 
One can see that the electromagnetic fields do not appear in this order and have no contribution to the total energy-momentum tensor in the first step.
And because the only non-zero source term of $h^{\alpha\beta}_{1\,\mathcal{W}}$, i.e., $T^{\alpha\beta}_{0\,\text{field}}$, have no portion in this order of the energy-momentum tensor, the effect of $h^{\alpha\beta}_{1\,\mathcal{W}}$ on $h^{\alpha\beta}_{1}$ is negligible and $h^{\alpha\beta}_{1}=h^{\alpha\beta}_{1\,\mathcal{N}}$.
In fact, each component of $h^{\alpha\beta}_1$ is only constructed from $T^{\alpha\beta}_{0\,\text{fluid}}$. So, we recover equations (7.37a)-(7.37c) of \cite{poisson2014gravity} for the components of $h^{\alpha\beta}_1$ and consequently equations (7.41a)-(7.41c) for the components of the metric in this step.

Our most important task at the second stage of the iterative procedure, as in the previous step, is to construct the source term of equation (\ref{general_int}) from the equations derived in the previous iteration.
As we know,  the required degree of accuracy to $\tau^{00}_1$, $\tau^{0j}_1$, and $\tau^{jk}_1$ is $O(1)$, $O(c)$, and $O(1)$, respectively.
With this in mind, the terms which do not survive at 1\tiny PN \normalsize approximation will be truncated, and therefore we will avoid additional calculations. 

The fluid contribution of the effective energy-momentum pseudotensor is similar to the single component perfect-fluid version introduced in section 7.3.1 of \cite{poisson2014gravity}. Therefore, we skip this part and focus our attention on the electromagnetic field contribution.

To find this contribution to the effective energy-momentum pseudotensor, i.e., $(-g_1)T^{\alpha\beta}_{1\,\text{field}}$, we first calculate the covariant components of $F^{\alpha\beta}_{1}$ by using $g_{\alpha\beta}^{1}$ and definition (\ref{E&B}). Then we substitute them
within (\ref{T_field}).
After truncating the result to the required order, we finally arrive at
\begin{eqnarray}
\label{T00_1field}
&& c^{-2}(-g_1)\,T^{00}_{1\,\text{field}}=\frac{1}{c^2}\frac{B^2}{2\mu_0}+O(c^{-4})\\
\label{T0j_1field}
&& c^{-1}(-g_1)\,T^{0j}_{1\,\text{field}}=O(c^{-2})\\
\label{Tjk_1field}
&&(-g_1)\,T^{jk}_{1\,\text{field}}= \frac{B^2}{2\mu_0}\delta^{jk}-\frac{B^jB^k}{\mu_0}+O(c^{-2})
\end{eqnarray}
in which $(-g_1)= 1+4U/c^2+O(c^{-4})$ is the metric determinant, $B^2/2\mu_0$ is the energy density of the magnetic field, $\epsilon_{M}$, and $B^2=\bm{B}.\bm{B}$.
Also, equation (\ref{Tjk_1field}) is reminiscent of the Maxwell stress tensor when there is no electric field in the system.
 
These equations tell us that the electromagnetic fields can play a significant role at this stage. More specifically, we straightforwardly see that the magnetic field appears in the gravitational field equations. On the other hand, there is no contribution from the electric field. This means that at the first PN limit, the electric field does not influence the curvature of spacetime. We postpone this very important discussion until subsection \ref{sub2_PN metric}, where we define the near-zone metric of the charged fluid in PN gravity.

To complete $\tau^{\alpha\beta}_1$, we need to calculate $\left(-g_1\right)t^{\alpha\beta}_{1\,\text{LL}}$ and $\left(-g_1\right)t^{\alpha\beta}_{1\,\text{H}}$. 
As we know,
these pseudotensors are constructed from $h^{\alpha\beta}_1$. Moreover, as we explained in the first iteration,
the electromagnetic fields do not appear in the components of $h_1^{\alpha\beta}$.
Therefore, 
$t^{\alpha\beta}_{1\,\text{LL}}$ and $t^{\alpha\beta}_{1\,\text{H}}$ turn out to be similar to the perfect-fluid ones, see chapter 7 in \cite{poisson2014gravity}.
It should be noted that, similar to the perfect-fluid system, the harmonic contribution has no effect on $\tau^{\alpha\beta}_{1}$, and can be discarded at this stage. 

Now, we collect all contributions, i.e., $ T^{\alpha\beta}_{1\,\text{fluid}} $, $ T^{\alpha\beta}_{1\,\text{field}} $, and $ t^{\alpha\beta}_{1\,\text{LL}} $, to obtain the PN expansion of the effective energy-momentum pseudotensor. Consequently, the source terms of the wave equation (\ref{wave_eq}) or equivalently the integrands in (\ref{general_int}), take the following forms.
\begin{eqnarray}
\label{tau00_1}
\frac{\tau^{00}_{1}}{c^{2}}&=&\rho^*+\frac{1}{c^2}\Big[\rho^*\Big(\frac{v^2}{2}-\frac{U}{2}+\Pi \Big) +\frac{B^2}{2\mu_0}\\\nonumber
&&-\frac{7}{16\pi G }\nabla^2U^2\Big]+O(c^{-4})
\end{eqnarray}
\begin{eqnarray}
\label{tau0j_1}
\frac{\tau^{0j}_{1}}{c}&=&\rho^*v^j+O(c^{-2})\\
\label{taujk_1}
\nonumber
\tau^{jk}_{1}&=& \rho^*\Big(v^jv^k-\frac{1}{2}U\delta^{jk}\Big)+\frac{B^2}{2\mu_0}\delta^{jk}-\frac{B^jB^k}{\mu_0}+p\,\delta^{jk}\\
&&-\frac{1}{16\pi G}\delta^{jk}\nabla^2U^2+\frac{1}{4\pi G}\partial^jU\partial^kU+O(c^{-2})
\end{eqnarray}
where $\rho^*=\rho\left(1+v^2/2c^2+3U/c^2\right)$
, and $U$ is the Newtonian gravitational potential defined in terms of $\rho^*$. 
Equations (\ref{tau00_1})-(\ref{taujk_1}) are our final results for the energy-momentum pseudotensor in the presence of the electromagnetic fields computed in the first PN approximation. Now we are ready to find the corresponding spacetime metric.

\subsection{Near-zone metric}\label{sub2_PN metric}
In this subsection, we first evaluate the components of the second-iterated gravitational potentials $h^{\alpha\beta}_2(x)$  in which $x$ is in the near-zone. 
In fact, they will be the basis of the near-zone metric in the second iteration.
Finally, we obtain the PN metric.
To do so, let us focus on the near-zone portion of the second-iterated potentials, $h^{\alpha\beta}_{2\,\mathcal{N}}$. 
At this stage, as we know, to derive the correct gravitational potentials, we need to use the gauge condition.
On the other words, $h^{\alpha\beta}_{2}$ must satisfy both equations (\ref{wave_eq}) and (\ref{conserv_eq}).
Therefore, here, we apply equations (7.15a)-(7.15c) in \cite{poisson2014gravity} (in which $\partial_{\beta}\tau^{\alpha\beta}=0$ has been used) for deriving the components of  $h^{\alpha\beta}_{2\,\mathcal{N}}$. After substituting equations (\ref{tau00_1})-(\ref{taujk_1}) into them, we recover equations (7.78a) and (7.78b) in \cite{poisson2014gravity} which contain an extra term $4\mathcal{B}/c^4$ related to the magnetic field, for the potentials $h^{00}_{2\,\mathcal{N}}$ and $h^{kk}_{2\,\mathcal{N}}$, respectively. Also, because the magnetic field has no contribution in the required order of $\tau^{0j}_{1}$, we can use equation (7.79) for $h^{0j}_{2\,\mathcal{N}}$.

The new potential $\mathcal{B}$ is defined as
\begin{equation}\label{B_int}
\mathcal{B}\left(t,\bm{x}\right)= G\int\frac{1}{2\,\mu_0}\frac{B'^2}{|\bm{x}-\bm{x}'|}d^3x'
\end{equation}
which is generated by the energy density of the magnetic field $\epsilon_{\text{M}}$ at time $t$ and position $\bm{x}'$. 
Similar to the definitions of the other PN potentials in the modern approach, the domain of this integrate is the three-dimensional surface $\mathcal{M}$ that is confined to the volume occupied by the charged matter distribution.
In fact, this equation can be thought of as a gravitational potential which its source term is the ``mass" of the magnetic field.
We henceforth refer to it as the \textit{gravitomagnetic potential}.

Before obtaining the near-zone metric, let us briefly consider one of the advantages of the modern approach as mentioned in Sec. \ref{introduction}.
As seen from this integration, it has non-compact piece which can manifestly extend over all space (i.e., the source term of the integral $\mathcal{B}$).
Therefore, it seems that this potential is ill-defined and this integral eventually diverges.
On the other hand, as we mentioned before, the domain of this integral is restricted to $\mathcal{M}$ in the modern approach. 
Consequently, the ambiguity in the definition of this type of potentials is eliminated and it is not divergent in this method.
While in the classic approach, the domain of the integration is not clearly specified and this potential has an ambiguous solution.
In fact, we are motivated by this advantage to utilize the modern approach as a result of which the gravitomagnetic potential is not ill-defined.

It is necessary to recall that, to complete the second-iterated potentials, we also have to calculate the wave-zone contributions to $h^{\alpha\beta}_{2}$.
In our system, in this step, $h^{\alpha\beta}_{2\,\mathcal{W}}$ is built from both $T^{\alpha\beta}_{1\,\text{field}}$ and $t^{\alpha\beta}_{1\,\text{LL}}$ which are non-compact sources and can undoubtedly exist beyond the near-zone.
Here, by using a crude estimation method, we show that this contribution to the potentials is negligible. Hence, this part of the potentials has no effect on the required order in the 1\tiny PN \normalsize context, and we are allowed to ignore $h^{\alpha\beta}_{2\,\mathcal{W}}$.  
Applying the source term $(-g_1)t^{\alpha\beta}_{1\,\text{LL}}$ constructed from $h^{\alpha\beta}_{1}$ in which the field point $\bm{x}$ is within the wave-zone, one can crudely estimate that this part of $h^{\alpha\beta}_{2\,\mathcal{W}}$ is of order $c^{-8}$.
To see the complete calculation of this estimate, we refer to \cite{poisson2014gravity}.
Here, we estimate the contribution arising from the existence of the magnetic field. More specifically, we estimate the least degree of accuracy of $h^{\alpha\beta}_{2\,\mathcal{W}}$ built from $T^{\alpha\beta}_{1\,\text{field}}$.
To do so, let us focus on the time-time component of $(-g_1)T^{\alpha\beta}_{1\,\text{field}}$ which is equal to $\epsilon_{\text{M}}=B^2/2\mu_0$.
It is important mentioning that the source functions inserted in the wave-zone integral (6.107) in \cite{poisson2014gravity} should have the specific form shown in equation (6.98).
As an example of a magnetic field with such a form, we examine a simple case where a moving point charge generates a magnetic field.
In this case, we assume that the charge and current densities are given by $\rho_{\text{e}}(\bm{x},t)=q\delta(\bm{x}-\bm{r}_0(t))$ and $\bm{J}_{\text{e}}(\bm{x},t)=\rho_{\text{e}}\bm{v}(t)$ respectively. 
Where $\bm{v}$ is the velocity of the charge $q$ and $\bm{r}_0(t)$ is its position at time $t$. 
The well-known expression for the magnetic field created by this source is (\citealp{jackson2007classical})
\begin{equation}\label{B_ret}
\bm{B}=\frac{\mu_0 q}{4\,\pi}\left\lbrace\frac{\bm{v}(\tau)\times\hat{\bm{R}}}{\kappa^2R^2}+\frac{1}{c\,R}\frac{\partial}{\partial t}\left[\frac{\bm{v}(\tau)\times \hat{\bm{R}}}{\kappa}\right]\right\rbrace
\end{equation}
where $\tau=t-R/c$ is the retarded time, $R=|\bm{x}-\bm{r}_0(\tau)|$ is the distance from the position of the point charge to the field point, and $\hat{\bm{R}}=(\bm{x}-\bm{r}_0(\tau))/R$ is a unit vector.
Also $\kappa=1-\bm{v}(\tau)\cdot\hat{\bm{R}}/c$ is a retarded factor.

As we know, this is the retarded solution for the magnetic field of a moving point charge.
Here, we assume that point charge is deeply within the near-zone, i.e., $\bm{r}_0(\tau)\ll\mathcal{R}$. 
And because $R$ is situated within the wave-zone, we find that $R\simeq|\bm{x}|=r$. 
Applying this assumption into equation (\ref{B_ret}) and after ignoring all numerical and angle-dependent factors, we obtain that the leading terms of $(-g_1)T^{00}_{1\,\text{field}}$ are given by
\begin{eqnarray}
\mu_0q^2 \Big(\frac{v^2(\tau)}{r^4}+\frac{2\,v(\tau)\dot{v}(\tau)}{c\, r^3}+\frac{\dot{v}^2(\tau)}{c^2r^2}+\dots\Big)
\end{eqnarray}
where an overdot displays differentiation with respect to $\tau$ and the ellipsis indicates higher order terms.
We can see that above terms have the form of equation (6.98) in which $n=4$ with $f(\tau)=\mu_0q^2\,v^2(\tau)$, $n=3$ with $f(\tau)=\mu_0q^2\,v(\tau)\dot{v}(\tau)/c$, and $n=2$ with $f(\tau)=\mu_0q^2\dot{v}^2(\tau)/c^{2}$.
If these expressions are substituted into the dominant term of equation (6.109), we find that $h^{00}_{2\,\mathcal{W}}$ built from $T^{00}_{1\,\text{field}}$ is of order $c^{-6}$.
By the same method, it can be shown that the time-space and space-space components of this part of the wave-zone potentials are of orders $c^{-5}$ and $c^{-6}$, respectively. Therefore, they explicitly play a role in the high order of PN expansion and have no effect on $h^{\alpha\beta}_2$ at 1\tiny PN \normalsize order. Consequently, we conclude that $h^{\alpha\beta}_2=h^{\alpha\beta}_{2\,\mathcal{N}}$.

We are now in a position to construct the PN metric. By substituting the components of $h^{\alpha\beta}_{2\,\mathcal{N}}$ which are introduced before, into (\ref{g00})-(\ref{det-g}) and keeping terms up to the required order, i.e., $g_{00}$ to order $c^{-4}$, $g_{0j}$ to order $c^{-3}$, and $g_{jk}$ to order $c^{-2}$, we find the following near-zone metric
\begin{eqnarray}
\label{g00_2}
&& g_{00}^2= -1+\frac{2}{c^2}U+\frac{2}{c^4}\left[\Psi+2\mathcal{B}-U^2\right]+O(c^{-6})\\
\label{g0j_2}
&& g_{0j}^2 = -\frac{4}{c^3}U^j+O(c^{-5})\\
\label{gjk_2}
&& g_{jk}^2 = \delta^{jk}(1+\frac{2}{c^2}U)+O(c^{-4})\\
\label{det-g2}
&&\left(-g^2\right)= 1+\frac{4}{c^2}U+O(c^{-4})
\end{eqnarray}
in which $\Psi=\psi+\partial_{tt}X/2$ .
Quantities $\psi$, $X$, and $\bm{U}$ are the well-known potentials in PN gravity, see chapter 7 of \cite{poisson2014gravity} for more details.
This is the spacetime metric of a charged system in PN gravity.   
In fact, it can approximately describe the true spacetime metric in the presence of the electromagnetic field to 1\tiny PN \normalsize order, while $\bm{x}$ is situated within the near-zone.
It should be recalled that this metric is made up of the potentials which are generated by the fluid and field variables.
Therefore, as expected, by ignoring the gravitomagnetic potential, i.e., discarding the field variable, the spacetime metric of the PN fluid in the harmonic gauge is recovered.

Now, to grasp the significant role of the magnetic field in the PN approximation, we incorporate $2\mathcal{B}$ within the definition of potential $\psi$, equation (7.73a) in \cite{poisson2014gravity}, where the pressure and internal energy of the fluid appear as source terms. So we define a new potential $\psi_{\text{M}}$ instead of $\psi$ as follows
\begin{eqnarray}\label{psi_B_int}
\psi_{\text{M}}\left(t,\bm{x}\right)= G\int\frac{{\rho^*}'\left(\frac{3}{2}v'^2-U'+\Pi_{\text{eff}}'\right)+3p_{\text{eff}}'}{|\bm{x}-\bm{x}'|}d^3x'
\end{eqnarray}
in which $\Pi_{\text{eff}}=\Pi+\Pi_{\text{M}}$ and $p_{\text{eff}}=p+p_{\text{M}}$ are the effective internal energy and pressure respectively.
Furthermore, $\Pi_{\text{M}}$ is defined as the magnetic energy density per unit mass of the fluid, i.e., $\Pi_{\text{M}}=\epsilon_{\text{M}}/\rho^*$,
and $p_{\text{M}}=\epsilon_{\text{M}}/3$ is 
the magnetic pressure.

This relation is reminiscent of the equation of state of radiation.
We see that the quantities $\Pi_{\text{M}}$ and $p_{\text{M}}$ are added to the internal energy and pressure of the fluid in equation (\ref{psi_B_int}), respectively.  
Therefore, one can conclude that in a charged system in the presence of the electromagnetic field, the pressure and internal energy should be modified to include the electromagnetic field contribution.
In fact, this modification in the potential $\psi$ is the main role of the electromagnetic field in the PN metric. 
Furthermore, as we already mentioned, only the magnetic field appears in the gravitational potentials and consequently in the metric of spacetime. This happens because the ratio of the electric energy density $\epsilon_0E^2/2$ to the magnetic energy density $B^2/2\mu_0$ is of order $c^{-2}$. Therefore, naturally, the electric field has no effect on the total $p$ and $\Pi$ at 1\tiny PN \normalsize context.
This fact illustrates that in addition to the usual internal energy and pressure of the fluid, the geometry of the spacetime depends on quantities related to the magnetic field in the first PN approximation.

It is also instructive to collect the small parameters by which we can characterize this PN system.
As we know, the PN approximation is based on the usual conditions in which 
\begin{eqnarray}\label{n3}
\frac{v^2}{c^2}\sim\frac{U}{c^2}\sim\frac{p}{\rho^*c^2}\sim\frac{\Pi}{c^2}\ll 1
\end{eqnarray}
Generalizing these assumptions for our PN system dealing with the magnetic field, we have
\begin{eqnarray}\label{n2}
\frac{\mathcal{B}}{c^2}\sim\frac{p_{\text{M}}}{\rho^*c^2}\sim\frac{\Pi_{\text{M}}}{c^2}\ll 1
\end{eqnarray}
This set of parameters mathematically indicates that the magnetic field should be weak to have a meaningful PN expansion for our system. These two sets of parameters characterize the charged PN fluid in the presence of the magnetic field. It is worth mentioning that conditions (\ref{n2}) directly mean that the magnetic energy density should be very small compared to the mass energy density, i.e., $\epsilon_{\text{M}}\ll \rho^*c^2$. Equivalently, a necessary condition for the charged fluid to lie in the first PN regime is that the magnetic field satisfies the 
condition $B\ll c\,\sqrt{2\mu_0 \rho^*}$ everywhere throughout the fluid. 
This also means that the Alfv\`{e}n velocity, i.e., $v_{\text{A}}=B/\sqrt{\mu_0 \rho^*}$, is small compared to the velocity of light.

\subsection{Post-Newtonian geodesic equation}\label{PN_geod}

In this subsection, we  
introduce the PN geodesic equation in the presence of the electromagnetic fields.
In fact, here, we study the impact of the electromagnetic fields on the geodesic equation of an uncharged particle which is freely falling under PN gravity.
We have shown in the previous subsection that even in 1\tiny PN \normalsize approximation, the PN metric depends 
on the magnetic variables (i.e., $p_{\text{M}}$ and $\Pi_{\text{M}}$).
Then the magnetic field has some gravitational effects in PN gravity that will be elegantly seen in the PN geodesic equation.

By using equations \eqref{g00_2}-\eqref{gjk_2} and corresponding Christoffel symbols\footnote{We should note that in the first PN limit, except for $\Gamma^{j}_{00}$, the magnetic field contribution does not appear in the other components of the Christoffel symbol. In $\Gamma^{j}_{00}$, $\psi$ is replaced with $\psi_{\text{M}}$. To see these components, we refer the reader to chapter 8 in \cite{poisson2014gravity}}, one can easily obtain
\begin{eqnarray}\label{PN_geo}
\nonumber
&&\frac{dv^j}{dt}=\partial_jU+\frac{1}{c^2}\Big[(v^2-4U)\partial_jU-(4v^k\partial_kU+3\partial_tU)v^j\\
&&-4v^k\left(\partial_jU_k-\partial_kU_j\right)+4\partial_tU_j+\partial_j\Psi_{\text{M}}\Big]+O(c^{-4})
\end{eqnarray}
where $\Psi_{\text{M}}=\psi_{\text{M}}+\partial_{tt}X/2$.
This is the PN version of the world line of an uncharged massive particle slowly moving in the presence of the electromagnetic field. This equation easily disentangles the gravitational effects of the electromagnetic fields from the pure electromagnetic forces.
Obviously, in the absence of an electromagnetic field, the last term on the right-hand side of equation (\ref{PN_geo}) coincides with $\Psi$, and the usual geodesic equation in PN theory is recovered. And also by dropping the terms of order $c^{-2}$, as expected, one can find the path of a free particle in Newtonian gravity.

\section{Magnetohydrodynamics in Post-Newtonian gravity}\label{MHD_PN gravity}
In this section, our main goal is to determine the PN version of Euler's equation given by the spatial component of $\nabla_{\beta}T^{\alpha\beta}=0$, and also the continuity equation. Furthermore, we find the PN version of Maxwell's equations. To find Euler's equation, we first construct each component of the total energy-momentum tensor from the PN metric to the required degree of accuracy.
In the subsection \ref{Euler_PN}, we will see that to derive Euler's equation to 1\tiny PN \normalsize order, we just need to know the correction terms of $O(1)$, $O(c^{-1})$, and $O(c^{-2})$ for $T^{00}$, $T^{0j}$, and $T^{jk}$ respectively.
For obtaining the components of $T^{\alpha\beta}$, let us build the covariant components of the electromagnetic field tensor from the PN metric as follows
\begin{eqnarray}
\label{covar_F0i}
F_{0i}&=&-\frac{E^i}{c}+\frac{4}{c^3}\left(\bm{U}\times\bm{B}\right)^i+O(c^{-5})\\
\label{covar_Fij}
F_{ij}&=&\epsilon_{ijk}B^k\left(1+\frac{4U}{c^2}\right)+O(c^{-4})
\end{eqnarray}

Now, by inserting equations (\ref{g00_2})-(\ref{gjk_2}) and (\ref{covar_F0i})-(\ref{covar_Fij}) within (\ref{T_fluid}) and (\ref{T_field}), and after combining the fluid and field contributions, we can obtain the components of $T^{\alpha\beta}$ as follows
\begin{eqnarray}
\label{T00_2}
T^{00}&=& c^2\rho^*\left[1+\frac{1}{c^2}\Big(\frac{v^2}{2}-U+\Pi\Big)\right]+\frac{B^2}{2\mu_0}+O(c^{-2})\\
\label{T0j_2}
T^{0j}&=& c\rho^*v^j\left[1+\frac{1}{c^2}\Big(\frac{v^2}{2}-U+\Pi+\frac{p}{\rho^*}\Big)\right]\\\nonumber
&&+\frac{1}{\mu_0 c}\left(\bm{E}\times\bm{B}\right)^j+O(c^{-3})\\
\label{Tjk_2}
T^{jk}&=& \rho^*v^jv^k\left[1+\frac{1}{c^2}\Big(\frac{v^2}{2}-U+\Pi+\frac{p}{\rho^*}\Big)\right]\\\nonumber
&&+p\Big(1-\frac{2U}{c^2}\Big)\delta^{jk}-\frac{1}{\mu_0c^2}\left[E^jE^k-\frac{1}{2}E^2\delta^{jk}\right]\\\nonumber
&&-\frac{1}{\mu_0}\left[B^jB^k-\frac{1}{2}B^2\delta^{jk}\right]\Big(1+\frac{2U}{c^2}\Big)+O(c^{-4})\\\nonumber
\end{eqnarray}
where $\left(\bm{E}\times \bm{B}\right)^j/\mu_0$ is the $j$-component of the Poynting vector that measures the flux of electromagnetic energy.

We next calculate the PN expansion of the continuity equation for the mass density $\rho^*$ and the PN Maxwell equations in subsections \ref{continuity_PN} and \ref{Maxwell_PN}, respectively. And finally, by using these equations, we simplify the PN Euler equation in subsection \ref{Euler_PN}.

\subsection{Post-Newtonian mass conservation}\label{continuity_PN}

Here, we briefly introduce the conservation of the mass density in PN gravity.
Mathematically, the conservation of the mass density expresses that $\nabla_{\mu}\left(\rho u^{\mu}\right)=0$. 
One can see that the magnetic field contribution does not appear in this relation. Therefore, the PN continuity equation of the mass density $\rho^*$ is recovered. So, we have

\begin{eqnarray}\label{conti_eq}
\frac{d\rho^*}{dt}=-\rho^*\nabla\cdot \bm{v}
\end{eqnarray}
in which $d/ dt=\partial/\partial t+\bm{v}\cdot\nabla$. 
In Newtonian gravity where there is no difference between $\rho^*$ and $\rho$, this relation reduces to the standard form of the continuity equation.

\subsection{Post-Newtonian Maxwell's equations}\label{Maxwell_PN}
To complete the governing equations describing the PN charged fluid, we must derive the PN version of Maxwell's equations.
To do so, let us start with the covariant form of Maxwell's equations

\begin{eqnarray}
\label{homo_eq}
&&\nabla_{\alpha}F_{\beta\gamma}+\nabla_{\gamma}F_{\alpha\beta}+\nabla_{\beta}F_{\gamma\alpha}= 0\\
\label{inhomo_eq}
&&\nabla_{\beta}F^{\alpha\beta}=\mu_0 J_{\text{e}}^{\alpha}
\end{eqnarray}
where $J_{\text{e}}^{\alpha}=\left(\rho_{\text{e}}c,\bm{J}_{\text{e}}\right)$ is the four-current density in which $\rho_{\text{e}}$ is the charge density and $\bm{J}_{\text{e}}$ is the three-current density of the fluid. Both quantities are measured in the laboratory frame.

We first focus on the homogeneous Maxwell equations.
As we know, equation \eqref{homo_eq} can be rewritten as $\partial_{\alpha}F_{\beta\gamma}+\partial_{\gamma}F_{\alpha\beta}+\partial_{\beta}F_{\gamma\alpha}= 0$. 
By setting $\alpha=i$, $\beta=j$, and $\gamma=k$ and substituting equation (\ref{covar_Fij}) into this relation, we get
\begin{equation}\label{div_B}
\nabla\cdot\bm{B}=-\frac{4}{c^2}\nabla U\cdot\bm{B}+O(c^{-4})
\end{equation}
after expanding and keeping terms up to order $c^{-2}$.
As seen from this equation, $\nabla\cdot\bm{B}\neq0$ in PN gravity. However, it should be noted that the $``\nabla"$ operator in this equation coincides with that of flat spacetime. Therefore, equation (\ref{div_B}) does not mean that there is magnetic monopole in the PN limit. 
In fact, it must be thought of as an unrealistic result of the choice of the flat-spacetime derivative operator in the weakly curved spacetime. 

So, completely similar to what we did for matter density $\rho^*$ to satisfy the convenient form of the continuity equation, we define a rescaled magnetic field $\bm{B}^*$ in terms of $\bm{B}$ and $U$ to hold the divergence-free form for the magnetic field. The result is
\begin{equation}\label{Bstar}
\bm{B}^*=\bm{B}\left(1+\frac{4\, U}{c^2}\right)
\end{equation}
for which we have
\begin{equation}\label{div_Bph}
\nabla\cdot\bm{B}^*=0
\end{equation}
This is one of the Maxwell equations in PN theory.
If we set $\alpha=i$, $\beta=j$, and $\gamma=0$ and then insert equations (\ref{covar_F0i}), (\ref{covar_Fij}), and \eqref{Bstar} into the homogeneous Maxwell equation, we find
\begin{eqnarray}\label{Fara_PN}
\frac{\partial\bm{B}^*}{\partial t}&=&-\nabla\times\bm{E}+\frac{4}{c^2}\Big\lbrace \nabla\times\left(\bm{U}\times\bm{B}^*\right)\Big\rbrace+O(c^{-4})
\end{eqnarray}

This is Faraday's induction equation in PN gravity. As can be seen clearly, by discarding the PN corrections, the standard form of Faraday's induction equation, i.e., $\partial\bm{B}/\partial t=-\nabla\times\bm{E}$, is recovered.

Here, we return to the inhomogeneous Maxwell equations, i.e., equation (\ref{inhomo_eq}), and calculate the PN expansion of these equations. One may easily rewrite equation (\ref{inhomo_eq}) as follows
\begin{equation}\label{inhomo_eq2}
\partial_{\beta}\left(\sqrt{-g}F^{\alpha\beta}\right)=\sqrt{-g}\mu_0 J_{\text{e}}^{\alpha}
\end{equation}
This equation simplifies to
\begin{equation}\label{div_E}
\nabla\cdot\bm{E}=\frac{\rho_{\text{e}}^*}{\epsilon_0}-\frac{2}{c^2}\Big\lbrace\frac{1}{\epsilon_0}\rho_{\text{e}}^*U+\nabla U\cdot\bm{E}\Big\rbrace +O(c^{-4})
\end{equation}
when we put $\alpha=0$ and apply equations (\ref{E&B}) and (\ref{det-g2}). $\epsilon_0$ is the dielectric constant and $\rho_{\text{e}}^*=\sqrt{-g}\rho_{\text{e}}$ is a rescaled charge density. At 1\tiny PN \normalsize order, this quantity is given by
\begin{equation}\label{rhoe_star}
\rho_{\text{e}}^*=\rho_{\text{e}}\left(1+\frac{2U}{c^2}\right)+O(c^{-4})
\end{equation} 

In fact, equation (\ref{div_E}) is the divergence equation of $\bm{E}$ in PN gravity. On the other words, this is the PN version of the Gauss's law to $O(c^{-2})$.
It is worth mentioning that equation (\ref{div_E}) can be represented in the traditional form by introducing a rescaled electric field $\bm{E}^*$. This field has the following form
\begin{equation}
\bm{E}^*=\bm{E}\left(1+\frac{2\,U}{c^2}\right)
\end{equation} 
by which we rewrite the PN divergence equation of $\bm{E}$ in terms of $\bm{E}^*$. Finally, in the PN framework, we obtain
\begin{equation}
\nabla\cdot\bm{E}^*=\frac{\rho^*_{\text{e}}}{\epsilon_0}
\end{equation} 
as the divergence equation of $\bm{E}^*$ in the flat-spacetime formulation. To find the last Maxwell equation, we need to study the spatial component of relation (\ref{inhomo_eq2}). In this case, after some manipulation, we obtain 
\begin{eqnarray}\label{Amper_PN}
\nonumber
\nabla\times\bm{B}^*&=&\mu_0\bm{J}_{\text{e}}^*+\frac{1}{c^2}\Big\lbrace \frac{\partial\bm{E}}{\partial t}+2\mu_0U\bm{J}_{\text{e}}^*\\
&&+2\nabla U\times\bm{B}^*\Big\rbrace+O(c^{-4})
\end{eqnarray}
where $\bm{J}_{\text{e}}^*=\sqrt{-g}\bm{J}_{\text{e}}$ is a rescaled three-current density.
Equation (\ref{Amper_PN}) is the PN version of Ampere's law. By discarding the gravitational fields, all of them are PN corrections, the standard form of Ampere's law with the displacement current, $\partial\bm{E}/\partial t c^2$, is recovered.
It should be noted that the displacement current due to the time-dependent field $E$ is of the same order of magnitude as the other PN corrections and, in principle, one may ignore it in the classical MHD regime where the electric field varies very slowly with time.

In fact, the PN Maxwell equations introduced above state what electromagnetic fields
are generated by the PN charged fluid.
It is also instructive to investigate the electromagnetic field equations by adopting the Lorenz gauge.

Here, we study electromagnetism in this gauge and introduce the corresponding wave equations in PN theory.
To do so, let us rewrite the electromagnetic field tensor $F_{\alpha\beta}$ as $F_{\alpha\beta}=\nabla_{\alpha}A_{\beta}-\nabla_{\beta}A_{\alpha}$, 
where $A^\alpha=\big(\frac{\phi}{c},\bm{A}\big)$ is the four-potential in which $\phi$ is the electrostatic potential; and $\bm{A}$ is the vector potential. In the following, we write the electric and magnetic fields in terms of $\phi$ and $\bm{A}$.
Inserting this definition into equation \eqref{inhomo_eq} and adopting the covariant form of the Lorenz gauge, i.e., $\nabla_\alpha A^\alpha=0$, the inhomogeneous Maxwell equations take the
simple form $\Box_{\text{g}}A^{\alpha}-R^{\alpha}_{\beta}A^{\beta}=-\mu_0J^{\alpha}_{\text{e}}$\footnote{For more detailed discussions on the derivation of this relation, we refer the reader to subsection 5.3.3 of \cite{poisson2014gravity}.},
where $\Box_{\text{g}}=g^{\alpha\beta}\nabla_{\alpha}\nabla_{\beta}=\nabla^{\beta}\nabla_{\beta}$ is the wave operator in the curved spacetime and $R_{\alpha\beta}$ is the Ricci tensor.
As we know, this is the curved-spacetime version of the electromagnetic wave equations in the Lorenz gauge.
One can rewrite this equation as follows
\begin{eqnarray}\label{EM wave2}
\nonumber
g^{\beta\gamma}&&\Big[\partial_{\beta}\partial_\gamma A^\alpha+\partial_{\beta}\Gamma^\alpha_{\eta\gamma}A^\eta+\Gamma^\alpha_{\eta\gamma}\partial_{\beta}A^\eta+\Gamma^\alpha_{\eta\beta}\big(\partial_\gamma A^\eta\\
&&+\Gamma^\eta_{\lambda\gamma}A^\lambda\big)
-\Gamma^\eta_{\gamma\beta}\big(\partial_\eta A^\alpha+\Gamma^\alpha_{\lambda\eta}A^\lambda\big)\Big]\\\nonumber
&&-g^{\alpha\gamma}R_{\gamma\beta}A^{\beta}=-\mu_0J^\alpha_{\text{e}}
\end{eqnarray}
By inserting the PN expansion of the Ricci tensor\footnote{In this system, the components of the Ricci tensor are similar to the perfect-fluid ones presented in chapter 8 in \cite{poisson2014gravity}. The only difference is that $\Psi$ is replaced with $\Psi_{\text{M}}$.} and the Christoffel symbols, and Setting $\alpha=0$ into equation \eqref{EM wave2}, we arrive at
\begin{eqnarray}\label{m90}
\nonumber
\Box \phi+\frac{2}{c^2}&&\Big\lbrace \nabla\cdot\left(\partial_tU \bm{A}\right)-\nabla^2\left(U \phi\right)+\nabla U\cdot\partial_t\bm{A}\\\nonumber
&&+\nabla\left(\nabla\cdot\bm{U}\right)\cdot\bm{A}+2\nabla^2\bm{U}\cdot\bm{A}-\nabla U\cdot\nabla\phi\\
&&+2\left(\partial_j U_k+\partial_k U_j\right)\partial_j A^k\Big\rbrace=-\frac{\rho_{\text{e}}}{\epsilon_0}
\end{eqnarray}
after expanding the solution up to $O(c^{-2})$ and some simplifications. 
In fact, this relation is the PN wave equation of $\phi$ in a weakly curved spacetime.
It should be recalled that $\Box=\eta^{\alpha \beta}\partial_{\alpha \beta}$ is the flat-spacetime wave operator. Similarly, by putting $\alpha=j$, the vector part of equation \eqref{EM wave2} takes the following form 
\begin{eqnarray}\label{m91}
\nonumber
\Box\bm{A}+\frac{2}{c^2}&&\Big\lbrace \nabla^2U \bm{A}-U\nabla^2\bm{A}+\left(\nabla U\cdot\nabla\right)\bm{A}+\nabla\bm{A}\cdot\nabla U\\
&&-\nabla U\left(\nabla\cdot\bm{A}\right)\Big\rbrace=-\mu_0\bm{J}_{\text{e}}
\end{eqnarray}
This is the PN version of the wave equation of the vector potential $\bm{A}$ expanded to 1\tiny PN \normalsize order. Equations \eqref{m90} and \eqref{m91} are the PN version of the electromagnetic wave equations under the Lorentz gauge.

We have not yet found the electric and magnetic fields in terms of $\phi$ and $\bm{A}$. To do so, let us return to the definition of the electromagnetic field tensor and use the simplified form of it as $F_{\alpha\beta}=\partial_\alpha A_\beta-\partial_\beta A_\alpha$. It is necessary to calculate the covariant components of the four-potential to the required PN order. One can easily build these components from the PN metric. In this case, the contravariant components of the four-potential $A^{\alpha}$ are given by
\begin{eqnarray}
\label{A_i}
&& A_i\simeq A^i+\frac{2U}{c^2}A^i+O(c^{-4})\\
\label{A_0}
&& A_0\simeq-\frac{\phi}{c}+\frac{1}{c^3}\Big(2U \phi-4U_j A^j\Big)+O(c^{-5})
\end{eqnarray}
Now, by inserting equation \eqref{A_i} into the definition of $F_{ij}$, and then comparing it with relation \eqref{covar_Fij}, we finally find
\begin{equation}\label{m93}
\bm{B}^*=\nabla\times\bm{A}+\frac{2}{c^2}\Big\lbrace U\nabla\times\bm{A}+\nabla U\times\bm{A}\Big\rbrace+O(c^{-4})
\end{equation}
as the definition of the magnetic field in terms of the vector potential containing PN corrections.
One can verify that this effective magnetic field is divergence free.
It is also instructive to introduce a rescaled vector potential $\bm{A}^*$ in order to abbreviate equation \eqref{m93}. So we have $\bm{B}^*=\nabla\times\bm{A}^*$
in which $\bm{A}^*=\big(1+\frac{2U}{c^2}\big)\bm{A}$. 

Substituting equations \eqref{A_i} and \eqref{A_0} into the definition of the time-space component of $F_{\alpha\beta}$ 
, and comparing the result with equation \eqref{covar_F0i}, we obtain
\begin{eqnarray}\label{m95}
\nonumber
\bm{E}=&&-\frac{\partial \bm{A}^*}{\partial t}-\nabla\phi^*-\frac{4}{c^2}\Big\lbrace\left(\bm{U}\cdot \nabla\right)\bm{A}^*+\left(\bm{A}^*\cdot\nabla\right)\bm{U}\\
&&+\bm{A}^*\times\left(\nabla\times\bm{U}\right)\Big\rbrace+O(c^{-4})
\end{eqnarray}
where $\phi^*=\phi\left(1-\frac{2U}{c^2}\right)$.
Equation \eqref{m95} expresses the electric field in terms of $A^*$, $\phi^*$, and $\bm{U}$ in PN gravity.

As a final remark in this subsection, we examine the charge conservation mathematically expressed by $\nabla_{\alpha}J^{\alpha}_{\text{e}}=0$, in PN gravity.
One can easily rewrite the charge conservation equation as $\partial_{\alpha}\left(\sqrt{-g}J^{\alpha}_{\text{e}}\right)=0$.
After applying the definition of $\Gamma^{\alpha}_{\beta\alpha}$ and substituting $\rho^*_{\text{e}}$ and $\bm{J}^*_{\text{e}}$, this equation becomes
\begin{eqnarray}\label{cons_charge2}
\frac{\partial \rho^*_{\text{e}}}{\partial t}+\nabla\cdot\bm{J}^*_{\text{e}}=0
\end{eqnarray}
Equation (\ref{cons_charge2}) is the conservation statement of charge in PN theory.

\subsection{Post-Newtonian Euler's equation}\label{Euler_PN}

One of the main equations to describe a fluid is the Euler or momentum equation. This equation simply relates the acceleration of a fluid element due to the underlying forces in the system. In our MHD system in the PN limit, we deal with the fluid and magnetic pressures as well as the electromagnetic and gravitational forces. Using the Euler equation we can directly measure the fluid element's response to these forces in PN framework. Therefore, let us study the relativistic generalization of the force equation to 1\tiny PN \normalsize order. As in the standard case, we focus on the conservation equation of the energy-momentum, i.e., $\nabla_{\mu}T^{\nu\mu}=0$, and then by using this equation, we proceed to extract the PN version of the Euler equation.

The conservation relation $\nabla_{\mu}T^{\nu\mu}=0$ can be written as
\begin{eqnarray}\label{EM_conser}
\partial_{\beta}\left(\sqrt{-g}T^{\alpha\beta}\right)+\Gamma^{\alpha}_{\beta\gamma}\left(\sqrt{-g}T^{\beta\gamma}\right)=0
\end{eqnarray}
Before moving on to derive Euler's equation, let us find $d\Pi/dt$ from the time component of equation (\ref{EM_conser}). In fact,  
we need $d\Pi/dt$ to order $c^{-1}$ in the calculations.
We construct $d\Pi/dt$ by substituting the Christoffel symbols and equations (\ref{T00_2})-(\ref{Tjk_2}) into the time component of equation (\ref{EM_conser}). 
By applying both the standard Euler equation of the MHD fluid 
and the standard Faraday equation as well as the vector identity $\nabla\cdot\left(\mathbf{a}\times \mathbf{b}\right)=\mathbf{b}\cdot\left(\nabla\times \mathbf{a}\right)-\mathbf{a}\cdot\left(\nabla\times \mathbf{b}\right)$, we arrive at
\begin{eqnarray}\label{eq_dPi2}
\nonumber
\rho^*\frac{d\Pi}{dt}&=&\frac{p}{\rho^*}\frac{d\rho^*}{dt}-\bm{v}\cdot\Big(\frac{1}{\mu_0}\left(\nabla\times\bm{B}\right)\times\bm{B}+\rho^*_{\text{e}}\bm{E}\Big)\\ &&+\frac{1}{\mu_0}\left(\nabla\times\bm{B}\right)\cdot\bm{E}+O(c^{-2})
\end{eqnarray}
This is the first law of thermodynamics for a charged fluid in the presence of the electromagnetic fields in Newtonian gravity.

Now, in order to find the PN version of Euler's equation, we set $\alpha=j$ and use the Christoffel symbols, the components of the energy-momentum tensor, and the continuity equation. Then, we insert equations (\ref{div_B}), (\ref{Fara_PN}), (\ref{div_E}), (\ref{Amper_PN}), and (\ref{eq_dPi2}) into the result. After applying the definition of $\bm{B}^*$ \footnote{To derive the shorter form of this relation, we do not make use of $\bm{E}^*$. Furthermore, in the following section, we will show that the electric force is of order $c^{-2}$. Consequently using $\bm{E}$ instead of $\bm{E}^*$ makes no difference in the final result.} and expanding the solution up to $O(c^{-2})$, we finally obtain
\begin{eqnarray}\label{Euler_PN2}
\nonumber
\rho^*\frac{d\bm{v}}{dt}&&=\rho^*\nabla U-\nabla p+\rho^*_{\text{e}}\bm{E}+\bm{J}^*_{\text{e}}\times\bm{B}^*+\frac{1}{c^2}\Big\lbrace\big(\frac{v^2}{2}+U\\\nonumber
&&+\Pi+\frac{p}{\rho^*}\big)\nabla p-\bm{v}\frac{\partial p}{\partial t}+\rho^*\Big[\left(v^2-4U\right)\nabla U\\\nonumber
&&-4\bm{v}\times\left(\nabla\times\bm{U}\right)+4\frac{\partial \bm{U}}{\partial t}-\bm{v}\big(3\frac{\partial U}{\partial t}+4\bm{v}\cdot\nabla U\big)\\\nonumber
&&+\nabla\Psi_{\text{M}}\Big]-\left(\bm{J}^*_{\text{e}}\cdot\bm{E}\right)\bm{v}
-\left(\rho^*_{\text{e}}\bm{E}+\bm{J}^*_{\text{e}}\times\bm{B}^*\right)\big(\frac{v^2}{2}\\
&&+3U+\Pi+\frac{p}{\rho^*}\big)\Big\rbrace+O(c^{-4})
\end{eqnarray}
This is the relativistic generalization of Euler's equation of the MHD fluid to order $c^{-2}$. If we ignore the electric and magnetic fields in the above equation, we recover the PN Euler equation of the perfect fluid which is first introduced in \cite{chandrasekhar1965post}. On the other hand, by discarding the PN corrections, its standard form for the MHD fluid is obtained.\footnote{It is necessary to mention that the electric force is of order $c^{-2}$, and consequently it should be neglected in the concept of the classical MHD regime.}

Note that in this section, we have derived the PN version of the MHD equations in the harmonic gauge. As mentioned before, these equations are directly obtained in the standard PN gauge in \cite{greenberg1971post}. Therefore, to compare our result with Greenberg's equations, we have to rewrite these MHD equations in the standard gauge of PN theory. In Appendix \ref{App_standard}, we introduce the relevant coordinate transformations \eqref{standard gauge} and the transformed quantities \eqref{transformed potentials}-\eqref{transformed quantities}, and then by applying them, we obtain the MHD equations under the standard PN gauge. As one may see in the Appendix \ref{App_standard}, there are some minor differences between our equations and those presented in \cite{greenberg1971post}.

It is instructive to recall that, as clarified in subsection \ref{PN_geod}, the gravitational effects of the electromagnetic field appear in $\nabla\Psi_{\text{M}}$ at 1\tiny PN \normalsize order.
According to this fact, the other terms in equation (\ref{Euler_PN2}) which depend on the electric and magnetic fields, are the relativistic contributions to the electromagnetic interaction in the system. In order to write these terms in a compact form, let us define a PN electromagnetic force per unit volume, $F_{\text{EM}}^*$, as $\bm{F}_{\text{EM}}^*=\bm{F}^*+\bm{F}_{\text{p}}$
in which
\begin{eqnarray}\label{Lorentz_PN}
\nonumber
\bm{F}^*&=&\big(\rho^*_{\text{e}}\bm{E}+\bm{J}^*_{\text{e}}\times\bm{B}^*\big)\Big(1-\frac{1}{c^2}\big[\frac{v^2}{2}+3U+\Pi\\
&&+\frac{p}{\rho^*}\big]\Big)+O(c^{-4})
\end{eqnarray}
is the PN version of the Lorentz force per unit volume, whose magnitude is clearly weaker than the standard one, and 
\begin{equation}\label{F-p}
\bm{F}_{\text{p}}=-\frac{1}{c^2}\left(\bm{J}^*_{\text{e}}\cdot\bm{E}\right)\bm{v}
\end{equation}
is the PN force created by the electromagnetic energy density $\bm{J}^*_{\text{e}}\cdot\bm{E}$ (\citealp{chiuderi2016basics}).
In fact, this force expresses that in addition to the Lorentz force, there is an extra rate of momentum transfer from the electromagnetic fields to the fluid element at 1\tiny PN \normalsize order.
So, as mentioned above, $\bm{F}_{\text{EM}}^*$ shows that the PN electromagnetic fields exert such PN electromagnetic force on the PN charged fluid elements. Moreover, equation (\ref{Euler_PN2}) reveals that in addition to the well-known PN forces which have been derived separately for the neutral perfect fluid in \cite{chandrasekhar1965post} and \cite{poisson2014gravity}, a charged particle moves under the influence of both PN electromagnetic forces and PN gravitational force of the electromagnetic fields in this case.

\section{The plasma equations in Post-Newtonian gravity}\label{PN plasma}

In this section, we assume that the plasma is a collisional one-fluid system in which the hydrodynamic equilibrium is established locally. 
Furthermore, to have the simplest description of the PN plasma, we apply the PN version of the MHD equations derived in the previous section. In other words, we attempt to study a weakly gravitating plasma obeying the first PN approximation, in the MHD regime.

Before moving on to derive the PN plasma equations, let us count the number of equations and fluid variables that characterize the MHD fluid in PN limit.
The equations governing a self-gravitating MHD fluid are given by equations (\ref{conti_eq}), (\ref{Fara_PN}), (\ref{Amper_PN}), (\ref{cons_charge2}), (\ref{eq_dPi2}), and (\ref{Euler_PN2}) and the PN Poisson equations of $U$, $\bm{U}$, $X$\footnote{i.e., equations (7.38), (7.39), and  (7.76) in \cite{poisson2014gravity}, respectively.}, and $\psi_{\text{M}}$, i.e., equation (\ref{psi_B_int}). Therefore, we have eighteen differential equations for twenty-two unknown variables $\rho^*, \bm{v}, p, \Pi, \bm{E}, \bm{B}^*, \bm{J}^*_{\text{e}}, \rho^*_{\text{e}}, U, \bm{U}, \psi_{\text{M}}, X$.
It is clear that to have a complete set of equations, we need four extra equations. To do so, it is common to take into account Ohm's law and the equation of state relating pressure $p$ to mass density $\rho^*$. 
In this case, by developing the PN version of this law and choosing an equation of state, we arrive at a complete set of governing equations which, in principle, are enough to investigate a weakly gravitating plasma. In the following subsection, we shall, therefore, introduce the relativistic generalization of Ohm's law to 1\tiny PN \normalsize order.

\subsection{Post-Newtonian Ohm's law}

In order to obtain the PN version of Ohm's law, as we did for other laws, we start with the covariant form of this law, and then keep the 1\tiny PN \normalsize corrections. Therefore, let us first mention the covariant form of  Ohm's law \citep{jackson2007classical} given as follows
\begin{eqnarray}\label{Ohm1}
J_{\text{e}}^{\mu}=\sigma F^{\mu\nu}u_{\nu}-\frac{1}{c^2}\left(J_{\text{e}}^{\nu}u_{\nu}\right)u^{\mu}
\end{eqnarray}
where $\sigma$ is the conductivity of the fluid and is related to the resistivity of the fluid, i.e., $\eta$, as $\eta=1/\sigma$. It should be noted that here, $\sigma$ is a scalar. As mentioned before, our aim is to study the quasi-neutral fluid (plasma) in PN gravity. So, taking into account this approximation where the fluid element is neutral in its comoving frame, we rewrite equation (\ref{Ohm1}) as the following relation
\begin{equation}\label{Ohm2}
J^{\mu}_e=\sigma F^{\mu\nu}u_{\nu}
\end{equation}
in which the Lorentz invariant $J^{\nu}_eu_{\nu}$ is zero. Now we set $\mu=i$ and use the PN expansion of 
$F^{\mu\nu}$ and $u_{\nu}$ to obtain the following expression
\begin{eqnarray}\label{J_e^*}
\nonumber
\eta\bm{J}^*_{\text{e}}&=&\bm{E}+\bm{v}\times\bm{B}^*+\frac{1}{c^2}\Big\lbrace\left(\bm{E}+\bm{v}\times\bm{B}^*\right)\big(\frac{v^2}{2}+U\big)\\
&&-4\,\bm{U}\times\bm{B}^*\Big\rbrace+O(c^{-4})
\end{eqnarray}
where we have used 
the definitions of $\bm{B}^*$ and $\bm{J}_{\text{e}}^*$
and expanded the solution up to order $c^{-2}$. As expected gravitational potentials appear in the PN version of Ohm's law.
Equation \eqref{J_e^*} provides three equations. So, these relations with an equation of state are sufficient to complete the previously mentioned set of the equations.
It is also instructive to investigate the time component of equation \eqref{Ohm2}.
Setting $\mu=0$, applying the same method as before, and using the definition (\ref{rhoe_star}) for $\rho_{\text{e}}^*$, we obtain 
\begin{equation}\label{rho_e^*}
\rho^*_{\text{e}}=\frac{\sigma}{c^2}\bm{E}\cdot\bm{v}+O(c^{-4})
\end{equation}
This is the rescaled charge density in terms of the electric field and fluid velocity in the laboratory frame. This relation reveals that although we assumed local quasi-neutrality, which implies that the fluid element is neutral in the rest frame, it possesses an electric charge in the laboratory frame even in the first PN limit.

\subsection{Post-Newtonian Plasma}\label{PN_plasma}
By introducing the PN version of Ohm's law, we are now in the position to complete the description of the PN plasma.
As the first step, let us eliminate the electric field from the (PN) MHD equations.
To do so, we find $\bm{E}$ from equation (\ref{J_e^*}), and after taking the curl of the result, we arrive at
\begin{eqnarray}\label{nabla*E}
\nonumber
\nabla\times\bm{E}&=&\eta\nabla\times\bm{J}_{\text{e}}^*-\nabla\times\left(\bm{v}\times\bm{B}^*\right)-\frac{1}{c^2}\Big\lbrace\eta\bm{J}_{\text{e}}^*\times\\
&&\big(\nabla U+\frac{1}{2}\nabla v^2\big)+\eta\nabla\times\bm{J}_{\text{e}}^*\big( U+\frac{v^2}{2}\big)\\\nonumber
&&-4\nabla\times\left(\bm{U}\times\bm{B}^*\right)\Big\rbrace+O(c^{-4})
\end{eqnarray}
Here, our next task is to remove $\bm{J}_{\text{e}}^*$ from equation \eqref{nabla*E}. To do so, we derive 
$\bm{J}_{\text{e}}^*$ from 
equation \eqref{Amper_PN}  
\begin{eqnarray}\label{J^*}
\nonumber
\bm{J}^*_{\text{e}}&=&\frac{1}{\mu_0}\nabla\times\bm{B}^*-\frac{1}{\mu_0 c^2}\Big\lbrace 2U\nabla\times\bm{B}^*+\frac{\partial}{\partial t}\big(\frac{\eta}{\mu_0}\nabla\times\bm{B}^*\\
&&-\bm{v}\times\bm{B}^*\big)+2\nabla U\times\bm{B}^*\Big\rbrace+O(c^{-4})
\end{eqnarray}
in which we use the standard form of Ohm's law, i.e., $\bm{E}=\eta\,\bm{J}^*_{\text{e}}-\bm{v}\times\bm{B}^*$, to eliminate the electric field. 
Using equation \eqref{J^*}, we replace $\bm{J}_{\text{e}}^*$ and also $\nabla\times\bm{J}_{\text{e}}^*$ into equation (\ref{nabla*E}).
By inserting the result
into relation (\ref{Fara_PN}), we finally arrive at  
\begin{eqnarray}\label{induc-eq-PN1}
\nonumber
\frac{\partial\bm{B}^*}{\partial t}&&=\frac{\eta}{\mu_0}\nabla^2\bm{B}^*+\nabla\times\left(\bm{v}\times\bm{B}^*\right)-\frac{\eta}{\mu_0c^2}\Big\lbrace \nabla^2\bm{B}^*\big( 3U\\\nonumber
&&+\frac{v^2}{2}\big)-2\nabla\times\big(\nabla U\times\bm{B}^*\big)-\left(\nabla\times\bm{B}^*\right)\times\big(3\nabla U\\\nonumber
&&+\frac{1}{2}\nabla v^2\big)+\frac{\partial}{\partial t}\Big(\frac{\eta}{\mu_0}\nabla^2\bm{B}^*+\nabla\times\big(\bm{v}\times\bm{B}^*\big)\Big)\Big\rbrace\\
&& +O(c^{-4})
\end{eqnarray}
as the magnetic induction equation with resistivity in PN gravity.
And this equation is clearly reduced to 
\begin{eqnarray}\label{induc-eq-PN}
\frac{\partial\bm{B}^*}{\partial t}=\nabla\times\left(\bm{v}\times\bm{B}^*\right)+O(c^{-4})
\end{eqnarray}
after ignoring the resistivity. This is the magnetic induction equation of ideal plasma in the first PN limit.

Now, we turn to the PN Euler equation and $\bm{E}$ in this equation.
It is important to keep in mind that $\rho_{\text{e}}^*$ is of order $c^{-2}$ when the fluid is quasi-neutral. 
So, in this case, we can rewrite the electric force as $\bm{F}_{\text{E}}=\left(\bm{J^*}_{\text{e}}\cdot\bm{v}\right)\bm{E}/c^2$,
after applying equation (\ref{rho_e^*}) and making use of the standard Ohm law. 
This equation implicitly expresses that the electric force can be completely ignored in the standard MHD equations in the Newtonian limit.
By adding equations (\ref{F-p}) to $\bm{F}_{\text{E}}$,
we obtain $\bm{F}_{\text{p}}+\bm{F}_{\text{E}}=\bm{J}^*_{\text{e}}\times\left(\bm{E}\times\bm{v}\right)/c^2$,
and then by substituting 
this relation within (\ref{Euler_PN2}), and again using the standard form of Ohm's law, we remove all electric fields from the PN Euler equation. So, equation (\ref{Euler_PN2}) becomes

\begin{eqnarray}\label{Euler_PN3}
\nonumber
\rho^*\frac{d\bm{v}}{dt}&&=\rho^*\nabla U-\nabla p+\bm{J}^*_{\text{e}}\times\bm{B}^*+\frac{1}{c^2}\Big\lbrace\big(\frac{v^2}{2}+U+\Pi\\\nonumber
&&+\frac{p}{\rho^*}\big)\nabla p-\bm{v}\frac{\partial p}{\partial t}+\rho^*\Big[\left(v^2-4U\right)\nabla U-\bm{v}\big(3\frac{\partial U}{\partial t}\\\nonumber
&&+4\bm{v}\cdot\nabla U\big)+4\frac{\partial \bm{U}}{\partial t}-4\bm{v}\times\left(\nabla\times\bm{U}\right)+\nabla\Psi_{\text{M}}\Big]\\
&&+\bm{J}^*_{\text{e}}\times\big(\eta\bm{J}^*_{\text{e}}\times\bm{v}-\left(\bm{v}\times\bm{B}^*\right)\times\bm{v}\big)\\\nonumber
&&-\bm{J}^*_{\text{e}}\times\bm{B}^*\big(\frac{v^2}{2}+3U+\Pi+\frac{p}{\rho^*}\big)\Big\rbrace+O(c^{-4})
\end{eqnarray}

Now let us rewrite the first law of thermodynamics given by equation \eqref{eq_dPi2}, and eliminate the electric field. Because the electric force is of order $c^{-2}$ in the quasi-neutral fluid, this term has no role in equation (\ref{eq_dPi2}). Notice that in equation \eqref{eq_dPi2} we keep terms up to $O(1)$. Therefore, we have   
\begin{equation}\label{PN energy-eq}
\rho^*\frac{d\Pi}{dt}=\frac{p}{\rho^*}\frac{d\rho^*}{dt}+\eta {J^*_{\text{e}}}^2+O(c^{-2})
\end{equation} 
after applying the standard form of Ampere's and Ohm's laws.



As a final task in this section, let us gather the MHD equations determining the behavior of PN plasma. As we have so far shown, the basic MHD equations that describe resistive plasma in the PN limit are equations (\ref{conti_eq}), (\ref{J^*}), (\ref{induc-eq-PN1}), (\ref{Euler_PN3}), and (\ref{PN energy-eq}).
This set of equations, together with the PN Poisson equations as well as the equation of state gives the complete description of a self-gravitating plasma in PN gravity.
By taking $\eta=0$, one can easily obtain the equations for an ideal plasma.
\section{summary and discussion}\label{summery}
In this paper, we have derived the MHD equations in PN gravity.
In Sec. \ref{Sec2}, first, we briefly introduced
the modern approach to PN theory. We then applied the modern PN procedure to obtain the spacetime metric of a charged system in the harmonic gauge.
We have assumed that the fluid part of our system is perfect.
Finally, we constructed the PN metric from
the PN gravitational potentials to 1\tiny PN \normalsize approximation.
We have shown that in addition to the well-known PN potentials, this metric is made up of the \textit{gravitomagnetic potential} $\mathcal{B}$. More specifically, this potential appears only in the time-time component of the metric. Furthermore, we have also shown that through the modern approach, there is no ambiguity in the definition of this new gravitational potential built from non-compact source.

In Sec. \ref{MHD_PN gravity}, we have demonstrated in detail how we can extract the PN version of the MHD equations by using the PN metric. In fact, we have found the relativistic corrections to the MHD equations to the required order $c^{-2}$ in PN gravity.
To do so, as a first step, we have obtained the components of the total energy-momentum tensor constructed from the field and fluid contributions.
Then we have introduced the PN mass conservation equation, the PN Maxwell equations, and the energy conservation equation.
Note that, just for the sake of completeness, we have also explored the PN version of the electromagnetic wave equations in the Lorenz gauge.
Finally, using these relations, we have simplified the spatial component of $\nabla_{\mu}T^{\nu\mu}=0$ and derived the relativistic generalization of Euler's equation of the MHD fluid to $O(c^{-2})$ in the PN approximation.

Also, we have recovered the MHD equations in the standard gauge of PN theory and then compared them with Greenberg's equations in Appendix \ref{App_standard}. To do so, we have briefly introduced the standard PN gauge transformation and the corresponding transformed potentials, electromagnetic fields, and other quantities; and then by utilizing them, we have found these equations in the standard gauge.

Moreover, we have defined two rescaled electric and magnetic fields, $E^*$ and $B^*$ respectively, to keep the traditional form of the divergence equations for these fields.
We have also expressed that the gravitational effects of the electromagnetic field appear in $\nabla\Psi_{\text{M}}$ term and the electromagnetic interaction is included in $F_{\text{EM}}^*$ term.
Indeed, we have introduced the PN electromagnetic force per unit volume containing the PN version of the Lorentz force and the PN force created by the electromagnetic energy density $\bm{J}^*_{\text{e}}\cdot\bm{E}$.
The MHD equations include the standard parts and some PN corrections. As expected, by ignoring the electromagnetic effects, these equations recover the governing equations of a neutral perfect fluid in the PN limit.

In order to make practical use of the MHD equations, we have studied the main equations determining the behavior of PN plasma in Sec. \ref{PN plasma}. Similar to the standard case, we have shown that it is necessary to introduce the PN version of Ohm's law for completing the required set of equations. 
By applying this useful equation, we have eliminated the electric field and charge density $\rho_{\text{e}}^*$ from the (PN) MHD equations. Indeed, we have reduced the number of unknown variables to eighteen. So we have had eighteen differential equations, by considering an equation of state, for eighteen unknown variables $\rho^*, \bm{v}, p, \Pi, \bm{B}^*, \bm{J}^*_{\text{e}}, U, \bm{U}, \psi_{\text{M}}, X$.
Using the PN Ohm law, we have also shown that the fluid element which is neutral in the rest frame possesses electric charge in the laboratory frame in the first PN limit.
Eventually, we have obtained the governing equations which can completely describe the resistive self-gravitating plasma in PN gravity. 
In fact, we have gathered the minimum number of equations by which we can completely describe the PN plasma in the MHD regime.

We should also mention that our analysis in this paper can be used as a first required step for higher-order PN derivations. In fact, to resolve some ambiguities in the foundation of the PN version of MHD equations already studied in the classic approach, we have used the modern approach to derive the PN version of MHD equations.
In this new method, the near-zone and wave-zone solutions of the wave form of the Einstein field equations are explicitly decomposed, and the PN metric is embedded in the near-zone.
Therefore, all PN potentials with non-compact sources which have the effective contribution of order $c^{-4}$ in the time-time component of the PN metric are completely devoid of any ambiguity.  
Moreover, one of the advantages of this method is that the retarded solutions are illustrated explicitly. So, especially in the higher orders of PN limit where these solutions, i.e., wave-zone potentials, have an important role, one can easily follow the footprint of the retarded effects in PN metric even in the presence of electromagnetic fields, for more detail, see subsection \ref{sub2_PN metric}.
To sum up, we have reformulated the PN version of the MHD equations presented in \cite{greenberg1971post} by applying the modern approach. This approach does not suffer from ambiguities in the classic approach, and can be used for future studies in the PN framework.

As we have already mentioned in the introduction section, the magnetic field can be super strong in some relativistic astrophysical systems, e.g., magnetars, AGN, GRBs, accretion disks, astrophysical jets and neutron stars. The magnetic field plays essential role in the dynamics of that systems. For example one of the main explanations for the formation and powering of astrophysical jets is the existence of strong magnetic fields (\citealp{blandford1977electromagnetic}). In these relativistic systems the nonlinear nature of general relativity combined with the complexities induced by the electromagnetic fields, prevent analytic descriptions. However, a useful first step toward the semi-analytic treatment of such systems can be achieved in the context of the PN MHD. 
So, we consider that the equations presented in this paper, as a first step, that would be helpful to study such relativistic systems.

Furthermore, the post-Newtonian MHD equations derived in this paper are useful to study linear instabilities of the relativistic fluid plasma environments. As another example, we mention the gravitational collapse in the presence of magnetic fields. This case has been widely investigated in the Newtonian regime, for example see \cite{nakano1998star,strittmatter1966gravitational,nakano1978gravitational,pudritz1985star}. Within the post-Newtonian MHD, it would be also interesting to investigate relativist corrections to the \textit{magneto-Jeans} instability (\citealp{elmegreen1987supercloud}).

Ultimately, we have introduced the first step to study these systems in a semi-analytic way. Of course, it is necessary to investigate higher order corrections, and especially the propagation of gravitational waves in these highly magnetized systems. In \cite{nazari2017post} and \cite{kazemi2018post}, we have used the first PN approximation to study the local gravitational stability of non-rotating and rotating neutral fluids, respectively. A similar analysis for MHD systems would be useful to obtain a better understanding of the stability issues in the highly magnetized relativistic systems (\citealp{nazari2018Jeans}).

\section*{acknowledgments}
This work is supported by Ferdowsi University of Mashhad under Grant No. 43364 (17/12/1395). We would like to thank Shahram Abbassi for continuous encouragement and support during this work. Also, helpful comments by the referee are gratefully acknowledged.

\bibliographystyle{apj}
\bibliography{short,PN_MHDv2}

\appendix

\section{Post-Newtonian magnetohydrodynamics in the standard gauge}\label{App_standard}

Here, we first introduce the standard gauge transformation of PN theory and then derive the PN version of MHD equations in this coordinate system, i.e., $(\bar{t},\bar{x}^{j})$. 
Finally, we compare these relations with the corresponding equations in \cite{greenberg1971post}, hereafter PG.

The following choice of coordinate transformation defines the standard gauge of PN theory (\citealp{poisson2014gravity})
\begin{eqnarray}\label{standard gauge}
t=\bar{t}+\frac{1}{2c^4}\partial_{\bar{t}}X+O(c^{-6})~~~~~~~,~~~~~~~x^{j}=\bar{x}^{j}+O(c^{-4})
\end{eqnarray}
Under this transformation, the 1\tiny PN \normalsize ordering of the metric is unchanged\footnote{Namely,  $O(c^{-4})$, $O(c^{-2})$ and $O(1)$ of the time-time component, $O(c^{-3})$ of the time-space component, and $O(c^{-2})$ and $O(1)$ of the space-space component of the PN metric are preserved.}
and the form of the PN metric does not change. Moreover, under this type of transformation, the PN gravitational potentials transform as follows
\begin{eqnarray}\label{transformed potentials}
\bar{U}=U~~~~,~~~~\bar{U}^j=U^j+\frac{1}{8}\partial_{\bar{t}\,\bar{j}}X~~~~,~~~~\bar{\Psi}_{\text{M}}=\Psi_{\text{M}}-\frac{1}{2}\partial_{\bar{t}\,\bar{t}}X
\end{eqnarray}
where the ``bar" potentials are transformed potentials in the new coordinate system; and the old potentials are written in terms of $\bar{t}$ and $\bar{x}^{j}$.
One can show that under the transformation \eqref{standard gauge}, the electric and magnetic fields change according to
\begin{eqnarray}\label{transformed fields}
&&\bar{\bm{E}}=\bm{E}+\frac{1}{2c^2}\frac{\partial}{\partial \bar{t}}\bar{\nabla}X\times\bm{B}+O(c^{-4})\\
\label{transformed fields2}
&&\bar{\bm{B}}=\bm{B}+O(c^{-4})
\end{eqnarray}
where $\bm{E}=\bm{E}(\bar{t},\bm{\bar{x}})$ and $\bm{B}=\bm{B}(\bar{t},\bm{\bar{x}})$; and the other quantities, e.g., $\rho^*$, $p$, and so on, do not change to the required PN order. In fact, these quantities transform as
\begin{equation}\label{transformed quantities}
\bar{Q}(\bar{t},\bar{x}^{j})\simeq Q(t,x^j)\simeq Q(\bar{t},\bar{x}^{j})+O(c^{-4})
\end{equation}

Now, by utilizing the coordinate transformation \eqref{standard gauge}, the corresponding transformed potentials \eqref{transformed potentials}, and the other transformed quantities \eqref{transformed fields}-\eqref{transformed quantities}, we find the PN version of MHD equations in the standard gauge. In the following, we gather these equations 
\begin{eqnarray}
&&\frac{d\rho^*}{d\bar{t}}=-\rho^*\bar{\nabla}\cdot \bm{v}\\
\label{energy eq}
&&\rho^*\frac{d\Pi}{d\bar{t}}=\frac{p}{\rho^*}\frac{d\rho^*}{d\bar{t}}+\eta {J^*_{\text{e}}}^2+O(c^{-2})\\
\nonumber
\label{Euler-eq}
&&\rho^*\frac{d\bm{v}}{d\bar{t}}=\rho^*\bar{\nabla} U-\bar{\nabla} p+\bm{J}^*_{\text{e}}\times\bm{B}^*+\frac{1}{c^2}\left\lbrace\big(\frac{1}{2}v^2+U+\Pi+\frac{p}{\rho^*}\big)\bar{\nabla} p-\bm{v}\frac{\partial p}{\partial \bar{t}}+\rho^*\Big[\left(v^2-4U\right)\bar{\nabla} U\right.\\\nonumber
&&~~~~~~~~~\left.-\bm{v}\left(3\frac{\partial U}{\partial \bar{t}}+4\bm{v}\cdot\bar{\nabla} U\right)+4\frac{\partial \bm{U}}{\partial \bar{t}}-4\bm{v}\times\left(\bar{\nabla}\times\bm{U}\right)+\bar{\nabla}\Psi_{\text{M}}\Big]-\left(\bm{J}^*_{\text{e}}\cdot\bm{E}\right)\bm{v}-\left(\bm{J}^*_{\text{e}}\times\bm{B}^*\right)\Big(\frac{1}{2}v^2+3U\right.\\
&&~~~~~~~~~\left.+\Pi+\frac{p}{\rho^*}\Big)+\bm{\sigma \bm{E}\big(\bm{E}\cdot\bm{v}\big)}\right\rbrace+O(c^{-4})\\
&&\bar{\nabla}\cdot\bm{E}=\frac{\rho_{\text{e}}^*}{\epsilon_0}-\frac{2}{c^2}\Big\lbrace\frac{\rho_{\text{e}}^*}{\epsilon_0}U+\bar{\nabla} U\cdot\bm{E}+\bm{\frac{1}{4}\bar{\nabla}\cdot\big(\frac{\partial}{\partial \bar{t}}\bar{\nabla}X\times\bm{B}^*\big)}\Big\rbrace +O(c^{-4})\label{nablaE2}\\
&&\bar{\nabla}\cdot\bm{B}^*= 0\\
\label{nablaE}
&&\bar{\nabla}\times\bm{E}=-\frac{\partial\bm{B}^*}{\partial \bar{t}}+\frac{4}{c^2}\Big\lbrace \bar{\nabla}\times\big(\bm{U}\times\bm{B}^*\big)\Big\rbrace+O(c^{-4})\\
&&\bar{\nabla}\times\bm{B}^*=\mu_0\bm{J}_{\text{e}}^*+\frac{1}{c^2}\Big\lbrace\frac{\partial\bm{E}}{\partial \bar{t}}+2\mu_0U\bm{J}_{\text{e}}^*+2\bar{\nabla} U\times\bm{B}^*\Big\rbrace+O(c^{-4})
\end{eqnarray}
in which $\bar{\nabla}$ is the gradient operator in the coordinates $\bar{x}^j$ and all the quantities are written in terms of $\bar{t}$ and $\bar{x}^{j}$. By comparing with equations \eqref{conti_eq}, \eqref{div_Bph}, \eqref{Fara_PN}, \eqref{div_E}, \eqref{Amper_PN}, \eqref{eq_dPi2}, and \eqref{Euler_PN2}\footnote{Hint: To have a better comparison, definitions \eqref{J_e^*} and \eqref{rho_e^*} should be inserted into the last two equations.}, we see that except one of the Maxwell equations, i.e., equation \eqref{nablaE2}, the rest of the above relations are invariant under this gauge transformation.

As we have pointed out in the introduction section, the PN version of MHD equations has been obtained under the standard gauge of PN theory in PG. Here, as mentioned above, we recover these equations by utilizing the modern approach of PN theory and then applying the standard gauge transformation.

In order to compare this set of equations with those presented in PG, we attempt to rewrite equations of PG in terms of our notation. It should be noted that the definition of the electric and magnetic fields are different in two methods. Here, let us express the fields presented in PG as $\bm{E}_{\text{G}}$ and $\bm{B}_{\text{G}}$. Now, we introduce the relationship between these fields and those introduced in our notation. Comparing equation (25) in PG with the corresponding equations, i.e., equations \eqref{covar_F0i} and \eqref{covar_Fij} written in the new coordinate system, in our method, we arrive at

\begin{eqnarray}
\label{E_G}
&&\bm{E}_{\text{G}}=\bm{E}-\frac{4}{c^2}(\bm{U}\times\bm{B}^*)\\
\label{B_G}
&&\bm{B}_{\text{G}}=\bm{B}^*
\end{eqnarray}
in which we utilize equations \eqref{transformed potentials}-\eqref{transformed fields2} and the definition of $\bm{B}^*$.

Similarly, we can rewrite the Maxwell eqautions introduced in PG in terms of our notation. To do so, let us start from quation (56) in PG, i.e., relations $\bar{\nabla}\times\bm{E}_{\text{G}}=-\partial\bm{B}_{\text{G}}/\partial\bar{t}$ and $\bar{\nabla}\cdot\bm{B}_{\text{G}}=0$. These relations take the following form
\begin{eqnarray}\label{eq3}
\bar{\nabla}\times\bm{E}=-\frac{\partial\bm{B}^*}{\partial \bar{t}}+\frac{4}{c^2} \bar{\nabla}\times\big(\bm{U}\times\bm{B}^*\big)
\end{eqnarray}
and 
\begin{eqnarray}\label{eq4}
\bar{\nabla}\cdot\bm{B}^*=0
\end{eqnarray}
respectively, after inserting definitions \eqref{E_G} and \eqref{B_G}.
Furthermore, by substituting Ohm's law, equation (53) in PG, and rewriting in SI units, one can show that equations (58) and (60) in PG can be expressed as 

\begin{eqnarray}\label{eq1}
\bar{\nabla}\cdot\Big\lbrace\bm{E}_{\text{G}}+\frac{1}{c^2}\Big[ 2U\bm{E}_{\text{G}}+\big(4\bm{U}-\frac{1}{2}\bar{\nabla}X_{\text{G}}\big)\times\bm{B}_{\text{G}}\Big]\Big\rbrace=\frac{\rho_{\text{e}}}{\epsilon_0}\Big(1+\frac{2U}{c^2}\Big)
\end{eqnarray}
and
\begin{eqnarray}\label{eq2}
\bar{\nabla}\times\Big[\Big(1-\frac{2U}{c^2}\Big)\bm{B}_{\text{G}}\Big]-\frac{1}{c^2}\frac{\partial\bm{E}_{\text{G}}}{\partial\bar{t}}=\mu_0\bm{J}_{\text{e}}\Big(1+\frac{2U}{c^2}\Big)
\end{eqnarray}
respectively, where $\bar{\nabla}X_{\text{G}}=-\bar{\nabla}X$. Applying the definition of $\bm{E}_{\text{G}}$, $\bm{B}_{\text{G}}$, $\rho_{\text{e}}^*$, and $\bm{J}_{\text{e}}^*$ and simplifying the results, we finally obtain

\begin{equation}\label{eq5}
\bar{\nabla}\cdot\bm{E}=\frac{\rho_{\text{e}}^*}{\epsilon_0}-\frac{2}{c^2}\Big\lbrace\frac{1}{\epsilon_0}\rho_{\text{e}}^*U+\bar{\nabla} U\cdot\bm{E}+\frac{1}{4}\bar{\nabla}\cdot\big(\frac{\partial}{\partial \bar{t}}\bar{\nabla}X\times\bm{B}^*\big)\Big\rbrace
\end{equation}
and
\begin{eqnarray}\label{eq6}
\bar{\nabla}\times\bm{B}^*=\mu_0\bm{J}_{\text{e}}^*+\frac{1}{c^2}\Big\lbrace\frac{\partial\bm{E}}{\partial \bar{t}}+2\mu_0U\bm{J}_{\text{e}}^*+2\bar{\nabla} U\times\bm{B}^*\Big\rbrace
\end{eqnarray}
for equations \eqref{eq1} and \eqref{eq2} in terms of our notation. As seen, equations \eqref{eq3}, \eqref{eq4}, \eqref{eq5}, and \eqref{eq6} are equal to the corresponding relations extracted from our calculation in the standard gauge of PN theory.

Here, we turn to the equation of energy conservation introduced in PG, i.e., equation (41). One can immediately realize that this relation can be exhibited as 
\begin{eqnarray}\label{eq7}
\rho^*\frac{d\Pi}{d\bar{t}}=\frac{p}{\rho^*}\frac{d\rho^*}{d\bar{t}}+\eta {J^*_{\text{e}}}^2+O(c^{-2})
\end{eqnarray}
after using equation \eqref{eq6} and the definition of $\rho^*$ and $J^*_{\text{e}}$ and then applying the first law of thermodynamics, namely
\begin{equation}
\nonumber
d\Pi+p\, d\Big(\frac{1}{\rho^*}\Big)=T_0 d S
\end{equation}
where $T_0$ is the local temperature and $S$ is the entropy density of fluid. We see that the rewritten form of equation (41) in PG, i.e., equation \eqref{eq7}, is equivalent to equation \eqref{energy eq}.

As a final task in this appendix, we obtain the Euler equation in PG in terms of our notation. To do so, we first rewrite the PN version of the Euler equation in PG, equation (69), and then by applying the Maxwell equations, we try to simplify the result. Finally, we compare it with the corresponding relation obtained in the standard gauge.

It is important to bear in mind that all of the gravitational potentials in PG, e.g., $U$, $\bm{U}$, $\Phi$, and so on, are considered as a function of $\rho$. In fact, to have a better comparison, it is necessary to rewrite these gravitational potentials in terms of $\rho^*$. So, with keeping this fact in mind, one can show that equation (69) reduces to 

\begin{eqnarray}
\nonumber
&&\rho^*\Big[1+\frac{1}{c^2}\big(\frac{1}{2}v^2+U+\Pi+\frac{p}{\rho^*}\big)\Big]\frac{d \bm{v}}{d \bar{t}}+\bar{\nabla}p -\rho^*\Big[1+\frac{1}{c^2}\big(\frac{3}{2}v^2-3U+\Pi+\frac{p}{\rho^*}\big)\Big]\bar{\nabla}U
-\frac{4\rho^*}{c^2}\Big(\frac{\partial \bm{U}}{\partial \bar{t}}-\bm{v}\times\big(\bar{\nabla}\times\bm{U}\big)\Big)\\\nonumber
&&-\frac{\rho^*}{c^2}\bar{\nabla}\Psi_{\text{M}}+\frac{\bm{v}}{c^2}\Big[\frac{\partial p}{\partial \bar{t}}+\bm{v}\cdot\bar{\nabla}p+\rho^*\bm{v}\cdot\frac{d\bm{v}}{d\bar{t}}+3\rho^*\frac{\partial U}{\partial \bar{t}}+3\rho^*\bm{v}\cdot\bar{\nabla}U\Big]-\frac{1}{\mu_0}\Big(1-\frac{6U}{c^2}\Big)\big(\bar{\nabla}\times\bm{B}_{\text{G}}\big)\times\bm{B}_{\text{G}}\\\nonumber
&&+\frac{1}{\mu_0c^2}\bm{v}\big[(\bm{E}_{\text{G}}+\bm{v}\times\bm{B}_{\text{G}})\cdot(\bar{\nabla}\times\bm{B}_{\text{G}})\big]-\frac{\bm{E}_{\text{G}}}{\mu_0 c^2}\bar{\nabla}\cdot\bm{E}_{\text{G}}+\frac{1}{\mu_0 c^2}\frac{\partial}{\partial \bar{t}}\big(\bm{E}_{\text{G}}\times\bm{B}_{\text{G}}\big)-\frac{1}{\mu_0 c^2}(\bar{\nabla}\times \bm{E}_{\text{G}})\times\bm{E}_{\text{G}}\\\nonumber
&&+\frac{2}{\mu_0 c^2}(\bar{\nabla}U\times\bm{B}_{\text{G}})\times\bm{B}_{\text{G}}=0
\end{eqnarray}
after utilizing the definition of $\rho^*$ and some manipulations. We recall that all of the gravitational potentials are written in terms of $\rho^*$ in the above equation. Using the Newtonian Euler equation along with equations \eqref{eq1}-\eqref{eq2}, and applying the simple vector identity $\mathbf{a}\cdot(\mathbf{b}\times\mathbf{c})=-\mathbf{b}\cdot(\mathbf{a}\times\mathbf{c})$, we obtain 
\begin{eqnarray}
\nonumber
&&\rho^*\frac{d\bm{v}}{d\bar{t}}=\rho^*\bar{\nabla} U-\bar{\nabla} p+\rho_{\text{e}}^*\bm{E}_{\text{G}}+\bm{J}^*_{\text{e}}\times\bm{B}_{\text{G}}+\frac{1}{c^2}\left\lbrace\big(\frac{1}{2}v^2+U+\Pi+\frac{p}{\rho^*}\big)\bar{\nabla} p-\bm{v}\frac{\partial p}{\partial \bar{t}}+\rho^*\Big[\left(v^2-4U\right)\bar{\nabla} U\right.\\\nonumber
&&~~~~~~~~~\left.-\bm{v}\left(3\frac{\partial U}{\partial \bar{t}}+4\bm{v}\cdot\bar{\nabla} U\right)+4\frac{\partial \bm{U}}{\partial \bar{t}}-4\bm{v}\times\left(\bar{\nabla}\times\bm{U}\right)+\bar{\nabla}\Psi_{\text{M}}\Big]-\left(\bm{J}^*_{\text{e}}\cdot\bm{E}_{\text{G}}\right)\bm{v}-\left(\bm{J}^*_{\text{e}}\times\bm{B}_{\text{G}}\right)\Big(\frac{1}{2}v^2+5U\right.\\\nonumber
&&~~~~~~~~~\left.+\Pi+\frac{p}{\rho^*}\Big)-2U\rho_{\text{e}}^*\bm{E}_{\text{G}}\right\rbrace+O(c^{-4})
\end{eqnarray}
in which we keep terms up to order $c^{-2}$.
Finally, by inserting definitions \eqref{E_G}, \eqref{B_G}, and \eqref{rho_e^*}, and expanding the result up to 1\tiny PN \normalsize order, we arrive at
\begin{eqnarray}
\nonumber
&&\rho^*\frac{d\bm{v}}{d\bar{t}}=\rho^*\bar{\nabla} U-\bar{\nabla} p+\bm{J}^*_{\text{e}}\times\bm{B}^*+\frac{1}{c^2}\left\lbrace\big(\frac{1}{2}v^2+U+\Pi+\frac{p}{\rho^*}\big)\bar{\nabla} p-\bm{v}\frac{\partial p}{\partial \bar{t}}+\rho^*\Big[\left(v^2-4U\right)\bar{\nabla} U\right.\\
&&~~~~~~~~~\left.-\bm{v}\left(3\frac{\partial U}{\partial \bar{t}}+4\bm{v}\cdot\bar{\nabla} U\right)+4\frac{\partial \bm{U}}{\partial \bar{t}}-4\bm{v}\times\left(\bar{\nabla}\times\bm{U}\right)+\bar{\nabla}\Psi_{\text{M}}\Big]-\left(\bm{J}^*_{\text{e}}\cdot\bm{E}\right)\bm{v}-\left(\bm{J}^*_{\text{e}}\times\bm{B}^*\right)\Big(\frac{1}{2}v^2+5U\right.\\\nonumber
&&~~~~~~~~~\left.+\Pi+\frac{p}{\rho^*}\Big)+\sigma \bm{E}\big(\bm{E}\cdot\bm{v}\big)\right\rbrace+O(c^{-4})
\end{eqnarray}

This equation is similar to equation \eqref{Euler-eq} with a minor difference in the numerical coefficient of term $\frac{U}{c^2}\bm{J}^*_{\text{e}}\times\bm{B}^*$.
As seen, in PG, the numerical coefficient of this term is $5$ while, in our derivation, the numerical coefficient of the corresponding term is $3$. Therefore, after a careful comparison, we could not confirm equation (69) in PG.

\end{document}